\newif\ifisonecolumn
\isonecolumntrue %use this to make it true (or comment if not)
 
\ifisonecolumn
\documentclass[onecolumn,draftcls,12pt]{IEEEtran}
\else
\documentclass{IEEEtran}
\fi
\ifisonecolumn
\newcommand{\TwoOneColumnAlternate}[2]{#2}
\else
\newcommand{\TwoOneColumnAlternate}[2]{#1}
\fi
\usepackage{etoolbox}

\newcommand{\FigDat}[2]{
\ifstrequal{#1}{System}{
\begin{figure}
\centering
{\includegraphics[width=#2]{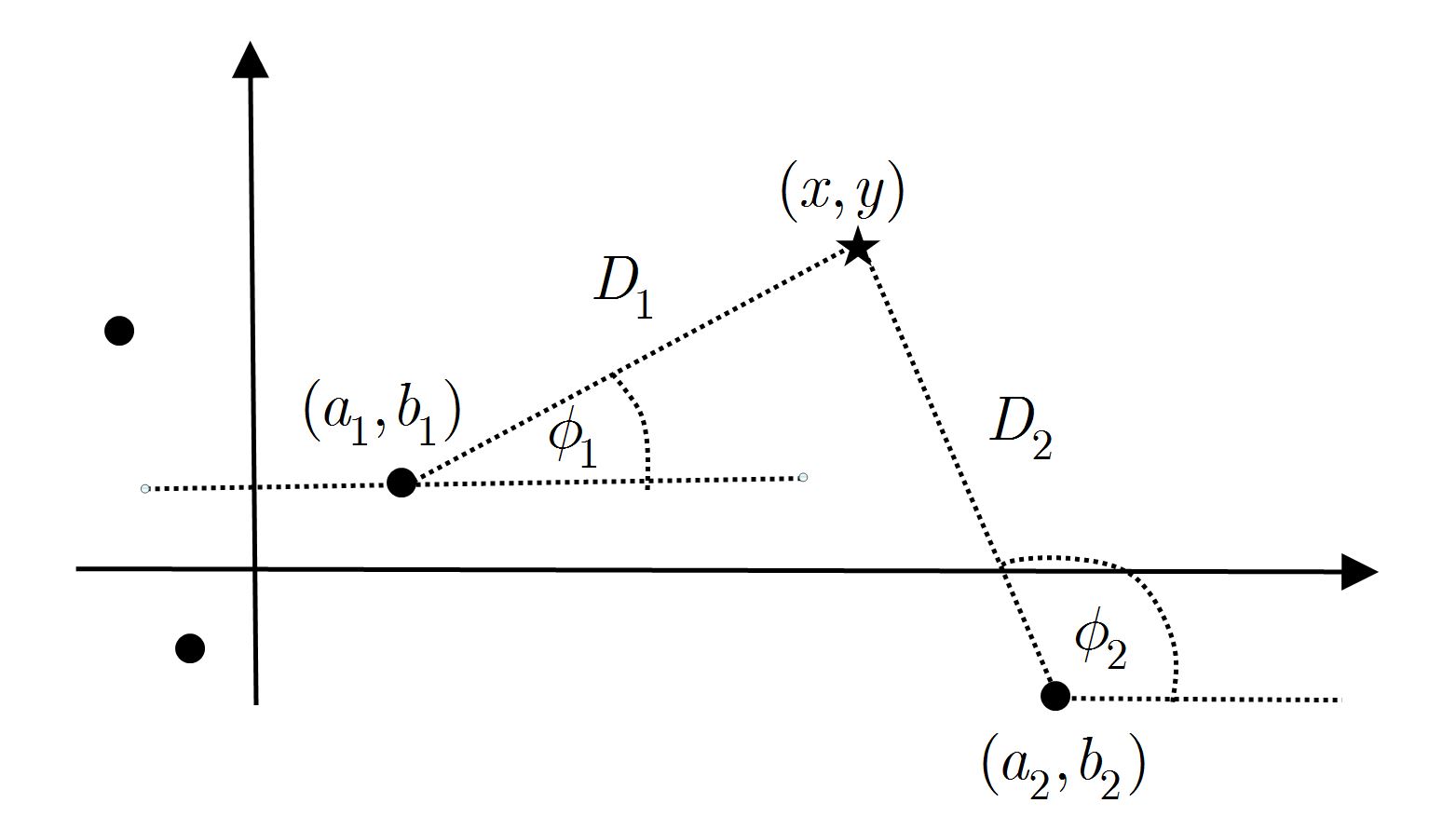}}
\caption{System model. The black circles denote the sensors;  the star denotes the source.  }
\label{Sensor2DPlain}
\end{figure} 
}{}
\ifstrequal{#1}{Performance}{
\begin{figure}[t]
        \includegraphics[scale=#2]{Fig_system}
    \caption{System model. .}
    \label{f:system model}
\end{figure}
}{}
}
\TwoOneColumnAlternate{
}{}

\usepackage{graphicx}
\usepackage{amsmath,amssymb,epsfig}
\usepackage{varioref}
\usepackage{subfigure}
\usepackage[sort&compress,square,numbers]{natbib}
\usepackage{blkarray}
\usepackage{commath}
\usepackage{arydshln}
\usepackage{enumerate}
\newcounter{my_set}
\newcounter{part}[my_set]

\newcommand{\part}[1]{\refstepcounter{part}{\noindent \bfseries
\itshape Part~\Alph{part}.}~\textit{#1}\\}
\newcommand{\arr}[1]{{{\mbox{$#1$}}}}
\newcommand{\Line}[2]{\TwoOneColumnAlternate{\\&&\hspace{#2}#1}{\\&&\hspace{#2}#1}}
\newcommand{\condnonumber}{\TwoOneColumnAlternate{\nonumber}{}}
\newcommand{\Brake}[1]{\TwoOneColumnAlternate{\nonumber\\&#1&}{#1}}
\newcommand{\BrakeL}[2]{\TwoOneColumnAlternate{\nonumber\\&&\hspace{#2}#1}{#1}}
\newcommand{\cond}[1]{\TwoOneColumnAlternate{#1}{}}
\newcommand{\Dcond}[2]{\TwoOneColumnAlternate{#1}{#2}}
\newenvironment{typew}{\sffamily}{}
\newcommand{\btw}{\begin{typew}}
\newcommand{\etw}{\end{typew}}
\newtheorem{theorem}{Theorem}
\newtheorem{proposition}[theorem]{Proposition}
\newtheorem{lemma}[theorem]{Lemma}
\newtheorem{corollary}[theorem]{Corollary}

\newtheorem{claim}[theorem]{Claim}

\newtheorem{definition}{Definition}
\newcommand{\ACRB}{\overline{\rm CRB}}

\newcommand{\beq}{\begin{equation}}
\newcommand{\eeq}{\end{equation}}
\newcommand{\beqI}{\begin{IEEEeqnarray}{rCl}}
\newcommand{\eeqI}{\end{IEEEeqnarray}}
\newcommand{\beqIl}{\begin{IEEEeqnarray}{lll}}
\newcommand{\eeqIl}{\end{IEEEeqnarray}}

\newcommand{\bea}{\begin{array}}
\newcommand{\ena}{\end{array}}
\newcommand{\bds}{\begin {itemize}}
\newcommand{\eds}{\end {itemize}}
\newcommand{\bdf}{\begin{definition}}
\newcommand{\blm}{\begin{lemma}}
\newcommand{\edf}{\end{definition}}
\newcommand{\elm}{\end{lemma}}
\newcommand{\bthm}{\begin{theorem}}
\newcommand{\ethm}{\end{theorem}}
\newcommand{\bprp}{\begin{prop}}
\newcommand{\eprp}{\end{prop}}
\newcommand{\bcl}{\begin{claim}}
\newcommand{\ecl}{\end{claim}}
\newcommand{\bcr}{\begin{coro}}
\newcommand{\ecr}{\end{coro}}
\newcommand{\bquest}{\begin{question}}
\newcommand{\equest}{\end{question}}

\newcommand{\xvec}{{\bf{x}}}

\newcommand{\zvec}{{\bf{z}}}

\newcommand{\ivec}{{\bf{i}}}

\newcommand{\Imat}{{\bf{I}}}

\newcommand{\Rmat}{{\bf{R}}}

\newcommand{\E}{{\rm{E}}}

\newcommand{\Tr}{{\rm Tr}}

\newcommand{\Mcal}{{\cal M}}
\newcommand{\fcal}{{\cal F}}
\newcommand{\Fcal}{{\cal F}}
\newcommand{\bcal}{{\cal B}}
\newcommand{\Bcal}{{\cal B}}

\newcommand{\real}{{\mathbb{R}}}
\newcommand{\sd}{{\mathbb{S}}}

\newcommand{\nat}{{\mathbb{N}}}

\newcommand{\intarr}{{\displaystyle\int}}

\newcommand{\define}{{\triangleq}}

\newcommand{\Psimat}{\mbox{\boldmath $\Psi$}}
\newcommand{\bPs}{\mbox{\boldmath $\Psi$}}

\def\btheta{{\mbox{\boldmath $\theta$}}}

\def\btheta{{\mbox{\boldmath $\theta$}}}

\def\thss{{\mbox{\scriptsize\boldmath $\theta$}}}

\def\psiss{{\mbox{\scriptsize\boldmath $\psi$}}}

\def\psivec{{\mbox{\boldmath $\psi$}}}
\def\bps{{\mbox{\boldmath $\psi$}}}
\def\bpsi{{\mbox{\boldmath $\psi$}}}
\def\psib{{\mbox{\boldmath $\psi$}}}

\def\th{\theta}

\def\thetavec{{\mbox{\boldmath $\theta$}}}
\def\bth{{\mbox{\boldmath $\theta$}}}
\def\thb{{\mbox{\boldmath $\theta$}}}

\def\ivec{{\mbox{\boldmath $i$}}}

\newcommand{\be}{\begin{equation}}
\newcommand{\ee}{\end{equation}}
\newcommand{\beqna}{\begin{eqnarray}}
\newcommand{\eeqna}{\end{eqnarray}}

\usepackage{color}

\begin{document}

\title{Lower Bound on the Localization Error in Infinite Networks with Random Sensor Locations}

\author{\IEEEauthorblockN{Itsik Bergel and Yair Noam  \thanks{ The authors are with the Faculty of Engineering,
Bar-Ilan University, Ramat Gan, Israel.
Email: \{itsik.bergel,yair.noam\}@biu.ac.il.
Some of the results in this paper were published \cite{7536794} in the IEEE 17th International Workshop on Signal Processing Advances in Wireless Communications (SPAWC), 2016.} }}

\maketitle

\date{}

\begin{abstract}
We present novel lower bounds on the mean square error (MSE) of the location estimation of an emitting source via a network where  the sensors are deployed randomly. The sensor locations are modeled as a homogenous Poisson point process. In contrast to previous bounds which are a function of the specific locations of all the sensors, we present CRB-type  bounds   on the expected   mean square error; that is, we first derive the CRB on the MSE as a function of the sensors' location, and then take  expectation  with respect to the distribution of the  sensors' location. Thus, these bounds are not a function of a particular sensor configuration, but rather of the sensor statistics.  Hence, these novel bounds can be evaluated prior to  sensor deployment and provide insights into design issues such as  the necessary  sensor density, the effect of the channel model, the effect of the signal power, and others.  The derived bounds are simple to evaluate and provide  a good prediction of the actual network performance.   

  \end{abstract}

\section{Introduction}
Localization via sensor networks   plays an important role in  many applications such as  personal security and safety, location based billing, machinery monitoring, radio-resource management,  intelligent
 transportation systems, etc. (e.g., \cite{paton1991terminal}\nocite{collier1994smart,stilp1996carrier,bussgang1998providing,koshima2000personal,tiwari2007energy}--\cite{polo2014semantic}).
To determine the source location,   each sensor takes measurements  about  the source; then, the  location is estimated by a joint processing of the measurements of all sensors. The sensors typically measure  time-of-arrival (TOA), time-difference-of-arrival (TDOA), angle-of-arrival (AOA), or received-signal-strength (RSS).

There has been extensive  research over the years  to improve  localization accuracy and robustness under various network conditions (see, for example, \cite{sun2005signal}\nocite{patwari2005locating,guvenc2009survey,wang2010survey}--\cite{han2013localization} and references therein).  A significant  part of  this  research  has to do with   localization performance analysis, which is carried
out primarily by deriving  the Cram\'er-Rao bound (CRB) on the mean square error (MSE) of the location estimate in  a variety of   network setups (e.g., \cite{patwari2003relative}\nocite{larsson2004cramer,qi2006time,fu2009cramer,sieskul2009hybrid}--\cite{closas2009cramer}).
In  all these works,   the derived  CRBs  are  functions of  the  sensor locations,   making it  less useful for designing  and  analyzing networks with arbitrary sensor deployment; e.g., if the sensors are dropped randomly from the air, or if  the designer is not familiar with the terrain  and/or the constraints under which the sensors will  actually be placed.   In such cases, a random network analysis is more suitable, where the sensors' location is modeled as a realization of a random  field. This approach  has  been applied to   different aspects of  ad-hoc wireless networks, and has  provided many important answers and insights,  including capacity scaling-laws  (e.g., \cite{tca2008}\nocite{tca2012}--\cite{DCM2012}), closed-form expressions for various performance metrics (e.g., \cite{apm2006}\nocite{arc2008,olt2009,aot2010}--\cite{haenggi2013local}), throughput bounds (e.g., \cite{tca2012},\cite{stamatiou2010channel}--\nocite{ERD2013}\cite{TEI2014}) and etc. 

In many respects, the problems of  ad-hoc networks and  localization are related; e.g.,  both share similar channel models, network structures and system parameters. Therefore, applying  stochastic geometry  to the analysis of source localization via sensor networks can lead to novel insights. Numerical studies of the localization  performance  of networks with random sensor placement   (e.g.; \cite{xi2010iterative}\nocite{zhou2010construction,li2014expectation}--\cite{naddafzadeh2014second})  have yielded  interesting results  about the behavior of random networks. An important approach  involves modeling the  sensor locations as Poisson Point Processes (PPP).  Lazos and Poovendran \cite{lazos2006hirloc} used PPP modeling in a numerical study of localization robustness. Aldalahmeh et al.   \cite{aldalahmeh2011distributed}    evaluated the probability of target detection by proximity sensors,  and  \cite{cevher2008pareto,cevher2009acoustic} further determined the scaling law of the localization error as a function of the sensor-density. Nevertheless,  to date,   no  closed-form expression   for the localization error  in random networks  has been put forward.

In this paper, using  stochastic geometry, we consider the sensor locations
  as  a realization of a homogenous PPP, and derive  CRB-type  bounds on the   achievable localization error. First, we derive the  CRB as a function of sensor
locations, in the case of an infinitely countable number of     sensors, each measuring a continuous time signal. While such a bound has been derived for a finite number of sensors, it has  never been properly extended to the case of an infinite countable number. This step, which also
includes the derivation of a likelihood
function, is crucial  to take the expectation with respect to the PPP distribution, and obtain a bound which is not a function of the sensors' location, but rather  a function of the sensor statistics.    We present a bound on the MSE in a network that utilizes both TOA and RSS measurements, and show that in the case of  joint TOA and RSS measurements, the bound  does not satisfy  the assumptions of \cite{cevher2008pareto} and does not scale as $\lambda^a$ for any $a>0$, where \(\lambda\) is the sensor density. We also provide closed-form asymptotic expressions for the bound in the wideband and narrowband cases. Note that the narrowband asymptote is equivalent to a network that only uses RSS measurements, while the wideband asymptote is equivalent to a network that only uses TOA measurements. Unlike the general bound, the two asymptotic bounds do scale as  $\lambda^a$ (such a  scaling was observed in \cite{cevher2008pareto}, using simplifying assumptions on the measurement model). Note that in this work we provide closed-form bounds, while the analysis of \cite{cevher2008pareto}  required a numerical evaluation of the  coefficients of this proportionality. 

 The remainder of the paper is organized as follows. In Section \ref{sec: system model} we present the system model, and in Section \ref{sec: CRB} we extend the known CRB for the case of a finite number of sensors  to the  case of an infinite number of sensors. In Section \ref{sec: Lower Bound} we present our novel lower bound on the localization error, and also give simple expressions for the bound in specific cases of interest. In Section \ref{NumericalResults} we present some numerical examples and Section \ref{sec:conclusions} concludes.

 Notation: Vectors and matrices  are denoted
by boldface lower-case and upper-case letters respectively.  \(\Vert\cdot\Vert\)
denotes the Euclidian norm. \(\real_+\)  denotes the positive
part of the real line.

\section{System Model} \label{sec: system model}
Consider a two-dimensional source localization
problem with an infinitely countable number of sensors, where sensor $m$ is located at the point $\psi_m=(a_m,b_m)\in\real^2 $, and the unknown source location
is \((x,y)\). We assume that the location of all sensors,     $\psivec=\{\psi_{m}\}_{m=1}^\infty,$   is a realization of a \(\lambda\)-density  homogenous PPP, \(\Psimat\), defined on the probability space \((\Omega,\fcal,P)\), and that  after the sensors have been randomly
 dispersed in the plane,  \(\psivec\) becomes known
to the fusion center. The  received signal at the  $m$-th sensor is given by\footnote{For simplicity, we limit the discussion to real signals.}: 
\beq \label{e: system model} r_{m}(t)= k_0D_{m}^{-\gamma/2}s(t-\tau_{m})+v_{m}(t), ~~~~t\in \real_{+},~m\in\nat, \eeq where \(s(t)\) is
a known signal waveform satisfying 
\begin{IEEEeqnarray}{rCl}\label{eq:RegularityConditionSquareIntegrableFirstAndSecondDerivative}
\left\vert \frac{d s(t)}{dt}\right\vert,\left\vert\frac{ d^{2} s(t)}{d t^{2}}\right\vert,\int_{0}^{\infty}\vert s^{}(t)\vert dt,\int_{0}^{\infty}\vert{d s(t)}/{d t}\vert
<\infty;
\end{IEEEeqnarray}
\(k_0\) is a constant  that depends    on the transmission power, antenna gains and carrier frequency. Furthermore,  $\gamma$ is the path-loss exponent\footnote{In this work we take the standard approach of assuming that the path loss exponent is known. See \cite{srinivasa2009path} for a discussion on the estimation of the path loss exponent in random networks.}, \(\tau_m=D_m/c\) is the time delay, $c$ is the speed of light,  
\beq\label{DefineDm} D_m(\thetavec)=\sqrt{(x-a_m)^{2}+(y-b_m)^{2}}\eeq is the distance between the source and the $m$-th sensor,  and  \(v_m(t)\) is
a white Gaussian noise with spectral density \(N_0/2\). 

The fusion center receives the signals from all of the  sensors, from which it  estimates  $\hat \thetavec=[\hat X,\hat Y]$. We denote the  estimation error by  $\Delta \bth \define \hat\bth-\bth$, and the mean square estimation error (MSE) for a given sensor  locations is  $\E\{\Vert \Delta\bth\Vert^2\vert\Psimat\}=\E\{(\hat X-x)^{2}+(\hat
Y-y)^2\vert\Psimat\}$. At the system design stage, the sensor locations are not yet known. Thus, in this work we focus on the expected MSE with respect to the distribution of the sensor locations: $\E\{\Vert \Delta\bth\Vert^2\}$.

\TwoOneColumnAlternate{
\FigDat{System}{80mm}
}{\FigDat{System}{120mm}
}

\section{Derivation of the CRB and average CRB}\label{sec: CRB}

 In this section we derive the average CRB by taking the expectation of the CRB over all possible sensor placements. 
To do so, we must first calculate the  CRB for a given sensor placement \(\psivec.\)  This CRB, which involves  an infinitely
 countable
 number of sensors has never been rigorously  derived. This derivation is mandatory
if we want to  take the expectation of the CRB with respect to the distribution of
\(\Psimat\) (see Appendix \ref{DerivationOfCRB} for further details).  

  \begin{theorem}\label{th: CRB def} 
Let \(\bpsi=\{\psi_{m}\}_{m=1}^{\infty}\in\real^{2\times\infty}\) be an infinite sequence of points in $\real^{2}$, where point $\psi_{m}=(a_{m},b_{m})$ denotes the location of the \(m\)th sensor, whose received signal is given in \eqref{e: system model}. For \(\Psimat=\psivec\), let $\hat \thetavec=[\hat X,\hat Y]$ be an unbiased estimate of the
source location \(\bth=[x,y]\in\real^{2};\) i.e., \(\E\{\hat\btheta-\bth\vert\psivec\}=0\).
If    \(\psib\) satisfies  \beq\label{FiniteSum} \sum_{m=1}^{
\infty}D_{m}^{-\gamma}<\infty,
\eeq
for every \(\thetavec\), where \(D_{m}\) is defined in \eqref{DefineDm},  then, the CRB on the mean square error
is:     
\begin{IEEEeqnarray}{rCl}
\E\{\Vert \hat\bth-\bth\Vert^2\vert\psivec\}\geq \mathrm{ CRB}(\thetavec,\bpsi),
\end{IEEEeqnarray}
where       
\begin{IEEEeqnarray}{rCl}
\label{e: CRB for finite}
\mathrm{ CRB}(\thetavec,\bpsi)=  \frac{1}{\rho} \frac{\sum _{m=1}^{\infty} g(D_m)}{ \sum _{m=1}^{\infty} \sum _{j > m} g(D_m)g(D_j)\sin ^2\left(\phi _m-\phi _j\right) },
\nonumber\\ \end{IEEEeqnarray}
\begin{IEEEeqnarray}{rCl}\label{d: g func}
g(D)=D^{-\gamma-2}\left( \gamma ^2+ {\frac{4 W_{e}}{c^2 }D^2} \right),
\end{IEEEeqnarray} \(E_{s}=k_{0}^{2}\int_{-\infty}^{\infty}s^{2}(t)dt\) is the received signal power at a unit distance,
$\rho=E_{s}/2N_{0}$, and  \( W_{e}=\frac{k_{0}^{2}}{E_{s}}\int_{-\infty}^{\infty}
\left({ds(t)}/{dt}\right)^{2}dt \)  is the effective bandwidth;   $\phi_m$  is the angle  between  the source and     the \(m\)th sensor; i.e.,  \(\cos \left(\phi _m\right)=\left(x-a _m\right)/D_m\) and \(\sin \left(\phi _m\right)=\left(y-b _m\right)/D_m\), as depicted in Fig. \ref{Sensor2DPlain}. \end{theorem}  \begin{IEEEproof}  
 The
CRB for  the model  \eqref{e: system model}
is already known in the case of a finite number of
sensors in deterministic locations \citep{sieskul2009hybrid}.
   However,  extending this result to the  infinite
case is not trivial and requires careful treatment. The full proof is given
 in  Appendix \ref{DerivationOfCRB}.
\end{IEEEproof}

The  bound  (\ref{e: CRB for finite}) is a function of the sensor locations,  \(\bpsi\) (due to the fact  that  \(\{\phi_{m}\}_{m=1}^{\infty}\) and \(\{D_{m}\}_{m=1}^{\infty}\) are functions of \(\psivec\)), which is assumed to be deterministic.  Next, by  treating the sensor locations as PPP \(\Psimat\) and taking the expectation of \(\mathrm{CRB}(\thetavec,\Psimat)\) with respect to the distribution of \(\Psimat\), we obtain  the average CRB.  This is a more   general tool that depends  on the sensor density, rather than a specific sensor locations.  Because   \(\mathrm{ CRB}(\bth,\bpsi)\) is defined solely under the constraint \eqref{FiniteSum}, one must  show that \(\mathrm{ CRB}(\bth,\bPs)\)  is a well-defined random variable. 
\begin{theorem}\label{Th:ExpectedCRBRaw}
 Let \(\Psimat\) be a PPP,  then \({\mathrm{ CRB}}(\bth,\Psimat)\) is a well-defined random variable; i.e., \eqref{FiniteSum} is satisfied with probability one.   \end{theorem}
\begin{IEEEproof} see Appendix \ref{ProofTheoremFiniteEnergy}.
\end{IEEEproof} 

Now that it is possible to take the expectation, the average CRB is given by 
\begin{IEEEeqnarray}{rCl}
\label{ExpectedCRB}
\ACRB(\bth)=\E\{{\rm CRB}(\bth,\Psimat)\}\TwoOneColumnAlternate{\\ \nonumber
&&\hspace{-35mm}}{}=\frac{1}{\rho}\E\left[\frac{\sum _m g(D_m)}{ \sum _m \sum _{j > m} g(D_m)g(D_j)\sin ^2\left(\phi _m-\phi _j\right) }\right].
\end{IEEEeqnarray}
It is important to stress that unlike \({\rm CRB}(\thetavec,\Psimat)\), \(\ACRB(\btheta)\) is not a bound on the estimation error that a particular network (with a particular \(\psivec\))  achieves, but rather a bound on the expected error  over all possible networks whose sensors' placement is a realization of a PPP.     

We conclude this section by a  discussion of the relationship between  \eqref{ExpectedCRB} and other CRB-type bounds  as well as the underlying unbias-condition required by each bound. The bound  \(\ACRB(\bth)\)  looks  similar to other   CRB-type bounds,  such as  the Hybrid CRB \citep{Schultheiss}, the modified CRB \citep{New_MMCRB_HAGIT,MMCRB_scalar,MMCRB_vector}  and the Miller-Chang bound \citep{modified_CRLB}, which were designed for   hybrid estimation problems; i.e.,  problems that  include  unknown  deterministic  parameters (for example, \(\bth\))   and    unknown random  (for example \(\bps\)) parameters.   Similar to \(\ACRB(\btheta)\), these bounds are obtained  by first deriving the CRB  (or the FIM in the modified CRB and the HCRB) for a given value of some random parameter, and then   taking the
expectation with respect to that random parameter.  However, the bound in   \eqref{ExpectedCRB} is  completely different, since the argument of the expectation here is   \({\rm CRB}(\thetavec,\psivec),\) which is derived under the assumption that  \(\bpsi\)    is  known to the estimator; i.e., it is not an unknown  parameter, or even an unknown  nuisance parameter.  With respect to the unbiased condition,  \(\ACRB(\bth)\)   requires that the  estimate of \(\thetavec\) be unbiased for every value of \(\psivec,\) except possibly  for sets of probability zero. This condition is similar (but
            not identical since in our case, \(\psivec\) is perfectly known) to that of Miller-Chang bound \citep{modified_CRLB}, and is stronger than the condition of the Hybrid and the Modified CRBs, which require unbiasedness only on the average. For further reference on CRB-type bounds, the reader is referred
to  \citep{notes_on_the_HCRB}.

\section{Lower bound on the localization error}\label{sec: Lower Bound}
Theorem \ref{Th:ExpectedCRBRaw} presents the average CRB on the localization error. This CRB can be  primarily  evaluated by Monte Carlo simulations (while truncating the sum over $m$ to a large enough number of sensors). The following theorem presents a simple lower bound on $\ACRB(\bth),$  expressed in terms of one-dimensional integrals. 
\begin{theorem}\label{th: main theorem}
Consider a sensor network  whose sensor locations is a realization of a homogenous PPP with  $\lambda$ sensors per unit area.  Then,  the average CRB in \eqref{ExpectedCRB} is lower bounded by:
\begin{IEEEeqnarray}{rCl}\label{e: theorem 1 bound} 
\ACRB(\thetavec) \ge {\rm CRB}_\mathrm{LB}
=\frac{ 4 }{\rho}\int_0^\infty \exp\left\{-2\pi\lambda Z(s)\right\}ds, 
\end{IEEEeqnarray}
where 
\begin{IEEEeqnarray}{rCl}\label{d: Z func}
Z(s)=\int_0^\infty\left[1-e^{-sg(r)}\right]r d r .
\end{IEEEeqnarray}
\end{theorem}

Although  presented in an integral form, the bound \({\rm CRB}_{\rm LB}\)  is  very informative. As expected,  the bound on the average CRB, \({\rm CRB}_{\rm LB}\)  scales  linearly  with the noise power and inverse linearly  with the transmission power. Moreover, because  $Z(s)$ is always positive,  \({\rm CRB}_{\rm LB}\) is a monotonically decreasing function of the sensor density, $\lambda$. 
\begin{IEEEproof} From   (\ref{ExpectedCRB})   
\begin{IEEEeqnarray}{lll}
\nonumber\ACRB(\thetavec)\Dcond{\\\nonumber&&\hspace{-11mm}=}{&=&}
\frac{1}{\rho}\E\left[\E\left[\left.\arr{ \frac{ \sum \limits _m g(D_m)}{ \frac{1}{2}\sum\limits _m \sum \limits _{j \ne m} g(D_m)g(D_j)\sin ^2\left(\phi _m-\phi _j \right)}  }\right|\{D_{m} \}_{m=1}^{\infty}\right]\right]
%\Line{\ge}{-11mm}
\Dcond{\\&&\hspace{-11mm}\geq}{\\&\geq&}
\label{SecondLine}\frac{1}{\rho} \E \left[ \arr{\frac{{\sum \limits_m g(D_m)}}{ \frac{1}{2}\sum\limits _m \sum\limits _{j \ne m} g(D_m)g(D_j)E[\sin ^2\left(\phi _m-\phi _j\right)\vert \{D_{m}\}_{m=1}^{\infty}] }}\right]
%\Line{=}{-11mm}
\Dcond{\\&&\hspace{-11mm}=}{\\&=&}
\label{ThirdLine}\frac{ 4 }{\rho}\E\left[\arr{\frac{\sum\limits _m g(D_m)}{ \sum\limits _m \sum\limits _{j \ne m} g(D_m)g(D_j) }}\right],
\end{IEEEeqnarray}
where  \eqref{SecondLine} follows from   Jensen's inequality for conditional expectations, and \eqref{ThirdLine}  is due to  the uniform distribution of the angles. Next, by adding  the missing diagonal terms of the double sum to the denominator, one obtains 
\begin{IEEEeqnarray}{rCl}
\ACRB(\thetavec)\nonumber
&\ge&\frac{ 4 }{\rho}\E\left[\frac{\sum _m g(D_m)}{ \sum _m \sum _{j \ne m} g(D_m)g(D_j)+\sum _m g^2(D_m) }\right] \\
&=&\frac{ 4 }{\rho}\E\left[\frac{\sum _m g(D_m)}{ \sum _m \sum _{j} g(D_m)g(D_j) }\right]
 \\\label{e: main bound derivation equation}
&=&\frac{ 4 }{\rho}\E\left[\frac{1}{ \sum _m g(D_m) }\right].
\end{IEEEeqnarray}

Considering the denominator in  (\ref{e: main bound derivation equation}), we define 
\begin{IEEEeqnarray}{rCl}\label{d: define G}
G=\sum _m g(D_m);
\end{IEEEeqnarray}
 a quantity whose  characteristic function is known in stochastic geometry  \cite{Venkataraman2006shot,baccelli2009stoc,haenggi2012stochastic}, 
\begin{IEEEeqnarray}{rCl}\label{e: G characteristic function}
\varphi_G(s)=\E\{e^{-sG}\}
=e^{-2\pi\lambda Z(s)}.
\end{IEEEeqnarray}
The average CRB can be evaluated from  \(\varphi_G(s)\)  via the formula (see for example \cite{stamatiou2010channel}):
\begin{IEEEeqnarray}{rCl}\label{e: expec one over x}
\E\{G^{-1}\}=\int_0^\infty \varphi_G(s)ds .
\end{IEEEeqnarray}
Substituting (\ref{e: G characteristic function}) and  (\ref{e: expec one over x}) into (\ref{e: main bound derivation equation}) establishes the desired result.
\end{IEEEproof}

As expected, \({\rm CRB}_{\rm LB}\),  scales  linearly  with the noise power and inverse linearly  with the transmission power. Moreover, because  $Z(s)$ is always positive,  \({\rm CRB}_{\rm LB}\) is a monotonically decreasing function of the sensor density, $\lambda$. However, obtaining further insights is more difficult because   the integrals in (\ref{e: theorem 1 bound}) and (\ref{d: Z func}) have no closed-form expression in  general.
To this end,  we derive closed-form expressions in  two extreme cases.

\subsubsection{Wideband extreme}
In many cases,  particularly in the case of wideband sources, the TOA  is  much more informative on the source's location  than the RSS is. Keeping that in mind, we now characterize the  CRB  in  the wideband extreme case. The intuition  is that  for large enough  $W_e$,   
\(
g(D)\approx  {4W_e D^{-\gamma }}/ {c^2},\) implying that  the RSS  is negligible. Substituting the latter approximation of  $g(D)$  into    (\ref{e: theorem 1 bound}) and (\ref{d: Z func}) yields:
\begin{IEEEeqnarray}{rCl}\label{e: wideband bound}
{\rm CRB}_\mathrm{LB,W}=\frac{c^2\left(\pi\lambda  \right)^{-\frac{\gamma}{2}}}{\rho W_e} \Gamma^{-\frac{\gamma}{2}}\left(1-\frac{2}{\gamma}\right) \Gamma\left(1+\frac{\gamma}{2}\right).
\end{IEEEeqnarray}
The following corollary formalizes the  approximation:
\begin{corollary}[Wideband localization]\label{cor: wideband} Consider 
 \({\rm CRB}_{\rm LB}\) of  Theorem
  \ref{th: main theorem} and \({\rm CRB}_{\rm LB,W}\) of \eqref{e: wideband bound}, then
\beq\label{Sandwits}\left(1-\frac{\pi\lambda c^2 \gamma   }{ 2W_{e}} \right){\rm CRB}_\mathrm{
 LB,W} < {\rm CRB}_{\rm LB}< {\rm CRB}_\mathrm{LB,W}; \eeq hence, ${\rm CRB}_{\rm LB}/ {\rm CRB}_\mathrm{LB,W}\longrightarrow1$ as \(W_{e}\longrightarrow\infty\) with a convergence rate of \(O(1/W_{e})\).
\end{corollary}
\begin{IEEEproof}
We first bound the integrand  (\ref{d: Z func}) from below and above, using
\beq\label{LongInEquality1}\begin{array}{lll}
1-\left(1-\gamma ^2  s r^{-\gamma -2}\right) e^{-4s W_{e} r^{-\gamma }/c^2}\Line{>}{-54mm}1-e^{-sg(r)}>1-e^{-4s W_{e} r^{-\gamma }/c^2}.
\end{array}\eeq
where for the right hand side (RHS) of (\ref{LongInEquality1}) we used the inequality \(x+1<e^x\) and  the left hand side (LHS) is because \(e^{-\gamma ^2 s r^{-(\gamma +2)}}<1\). Let 
\beq
U(s)\triangleq\int_{0}^\infty(1-e^{-4s W_{e} r^{-\gamma }/c^2})rdr
\eeq
Then, from (\ref{LongInEquality1})
\begin{IEEEeqnarray}{lll}\label{ThreeInequality}
U(s)<Z(s)&&<\int_0^{\infty } \arr{\big(1-\big(1-\gamma ^2 s r^{-\gamma -2}\big) e^{-4s W_{e} r^{-\gamma }/c^2})rdr}\condnonumber
\Line{=}{00mm}U(s) +\frac{ c^2\gamma   }{ 4W_{e}}
\end{IEEEeqnarray}
  where for the RHS we used the  identity
\begin{IEEEeqnarray}{rCl}\label{e: simple gamma integral formula}
\int_0^\infty e^{-bx^{-a}}x^{-1-c}dx= 
\frac{b^{-\frac{c}{a}}}{a}\Gamma\left(\frac{c}{a}\right).
\end{IEEEeqnarray}
From \eqref{ThreeInequality}
\begin{IEEEeqnarray}{lll}
 {\rm CRB}_\mathrm{LB,W} =\frac{ 4 }{\rho}\int_0^\infty e^{-2\pi\lambda U(s)}ds<{\rm CRB}_{\rm LB}\\< \frac{ 4 }{\rho} \int_0^\infty e^{-2\pi\lambda\left(U(s)+  \frac{c^2\gamma}
 {4W_{e}}\right)}ds<\arr{\left(1-\frac{ 
\pi\lambda  c^2\gamma}{ 2W_{e}}\right)}{\rm CRB}_\mathrm{LB,W}\nonumber
\end{IEEEeqnarray}
where we used   $\exp\{-x\}>1-x$. This  establishes the proof.
\end{IEEEproof}

The inverse linear relation between \({\rm CRB}_{\rm LB}\) in (\ref{e: wideband bound}) and the effective bandwidth is, again, quite expected. However, this bound also indicates the  effect of the sensor-density and  the path-loss exponent; i.e.,  it decreases as $\lambda^{-\gamma/2}$. Thus, for  $\gamma$ which are close to 2, the bound is inversely proportional to the sensor-density.  In this extreme case, the sensor density \(\lambda\) has a similar effect as the number of sensors in the more traditional setup of sensors with independent observations. On the other hand, in the more typical setups where $\gamma>2,$ the density \(\lambda\) has a stronger effect than the number of sensors. This may be explained by the fact that large values of \(\lambda\) indicate a high probability of  having  sensors  in close proximity to the target. 

 \subsubsection{Narrowband extreme}
Here,  the TOA is not informative enough and the localization relies primarily
 on the RSS information. One practical example (that is not covered by this model) is the case of a dense multipath, where it is very difficult to determine the location from TOA information. Another example, which is easily obtained from our model, is the narrowband extreme case, where $W_e$ is small enough such  that the TOA data is insignificant; i.e.,  
\(g(D) \approx\gamma ^2 D^{-\gamma -2}.\)
Again, similar to \eqref{e: wideband bound} the approximate bound is given by
\begin{IEEEeqnarray}{rCl}\label{e: NArrowband bound}
{\rm CRB}_\mathrm{LB,N}=\frac{ 4 }{\rho\gamma ^2  }&\left(\pi\lambda  \Gamma\left(\frac{\gamma}{\gamma+2}\right) \right)^{-\gamma/2-1}\Gamma\left(2+\frac{\gamma}{2}\right).\;\;\;\; \end{IEEEeqnarray}
The exact formulation is established in the following corollary.
\begin{corollary}[Narrowband localization]\label{cor: narrowband}
The CRB bound of Theorem \ref{th: main theorem} satisfies
\begin{IEEEeqnarray}{rCl}\label{NB-Extrerme}
\vert {\rm CRB}_\mathrm{LB,N }-{\rm CRB}_{\rm LB}\vert<O\left(W_{e}\right), 
 \end{IEEEeqnarray}
where \({\rm CRB}_\mathrm{LB,N}\) is
given in \eqref{e: NArrowband bound}.  Hence, ${\rm CRB}_{\rm LB}$ converges to ${\rm CRB}_{LB,N}$ as the bandwidth decreases.

\end{corollary}
\begin{IEEEproof} 
Again, we bound the integrand  (\ref{d: Z func}) from below and above,\beq\label{LongInEquality}
1-e^{-s\gamma ^2  r^{-\gamma -2}}
 \left(1-\arr{\frac{4 sW_{e} r^{-\gamma }}{c^2}}\right)>1-e^{-s g(r)}>1-e^{-s \gamma ^2   r^{-\gamma -2}}.
\eeq
Similar to \eqref{LongInEquality1}, integrating the RHS of \eqref{LongInEquality} yields ${\rm CRB}_{\rm LB}< {\rm CRB}_{\rm LB,N}$,
which establishes the RHS of \eqref{NB-Extrerme}.
To show the LHS of \eqref{NB-Extrerme},
we  multiply the LHS  of \eqref{LongInEquality} by $r$, integrate  and by using (\ref{e: simple gamma integral formula})
we obtain
\begin{IEEEeqnarray}{rCl}
Z(s)&<&\int_0^{\infty } 
\arr{\left(1-e^{-s\gamma ^2   r^{-\gamma -2}}
 \left(1-\frac{4 sW_{e} r^{-\gamma }}{c^2}\right)\right)rdr}\Brake{=} V(s)+Q(s) 
\end{IEEEeqnarray}
where 
\beq
\begin{array}{ll} V(s)\triangleq
{\displaystyle\int}_0^{\infty } \left(1-e^{-s \gamma ^2   r^{-\gamma -2}}
 \right)r dr\\[1em]
Q(s)\triangleq\frac{4 sW_{e}(s \gamma ^2 )^{\frac{1-\gamma}{\gamma + 2}}}{c^2(\gamma +2)}\Gamma\left(\frac{\gamma-1}{\gamma +2}\right).
\end{array}
\eeq
Hence 
\beq
\begin{array}{lll}
{\rm CRB}_{\rm LB,N}=\frac{ 4 }{\rho}{\displaystyle\int}_0^\infty \exp\left\{-2\pi\lambda V(s)\right\}ds\\[1em]~~~~~~~~~~~~>{\rm CRB}_\mathrm{LB}
=\frac{ 4 }{\rho}{\displaystyle\int}_0^\infty \exp\left\{-2\pi\lambda Z(s)\right\}ds\\[1em]~~~~~~~~~~~~>\frac{ 4 }{\rho}{\displaystyle\int}_0^\infty \exp\left\{-2\pi\lambda (V(s)+Q(s))\right\}ds
\end{array}
\eeq
and therefore
\begin{IEEEeqnarray}{rCl}
\vert{\rm CRB}_{\rm LB}&-&{\rm CRB}_\mathrm{LB,N }\vert\BrakeL{<}{-15mm} \frac{ 4 }{\rho}\int_0^\infty \big(e^{-2\pi\lambda V(s)}-e^{-2\pi\lambda(V(s)+Q(s))}
  \big)ds\\&&\hspace{-15mm}<\frac{ 4 }{\rho}\int_0^\infty e^{-2\pi\lambda
V(s)}
\big(1 -e^{-2\pi\lambda Q(s)} \big)ds  \BrakeL{<}{-15mm}\frac{ 4 }{\rho}\int_0^\infty e^{-2\pi \lambda 
 V(s)}
 2\pi \lambda Q(s) ds \Dcond{\BrakeL{<}{-15mm}}{\\&&<} \frac{16 \pi ^{-\frac{1}{2} (\gamma +3)} \lambda  W_{e}}{c^2 \gamma ^4 \rho }\begin{array}{lll}\Gamma\left(\frac{\gamma}{\gamma +2}\right)^{-\frac{\gamma+5 }{2}}\Gamma \left(\frac{\gamma -1}{\gamma +2}\right)\Gamma\left(\frac{\gamma+5 }{2}\right)\end{array}\nonumber \end{IEEEeqnarray}
where in the third line we used   \(e^{-x}>1-x\) for \(x>-1\). This establishes the proof. 
\end{IEEEproof}
The performance in the narrowband case is typically  not as good  as in the wideband case, due to the effective bandwidth factor, $W_e$, in (\ref{e: wideband bound}) which significantly reduces the bound. Corollary \ref{cor: narrowband} shows that in the narrowband regime, the bound decreases  as $\lambda^{-\gamma/2-1}$,   which is a faster decay rate than in the wideband case. This dependence shows that RSS localization depends heavily  on the probability of having sensors very close to the target.

\section{Numerical examples}
\label{NumericalResults}
\begin{figure}[t]
\centering
%{\includegraphics[width=75mm]{CRB_wext_bound_vs_Es_final.eps}}
\includegraphics[width=75mm,trim={0 0 1cm 15cm},clip]{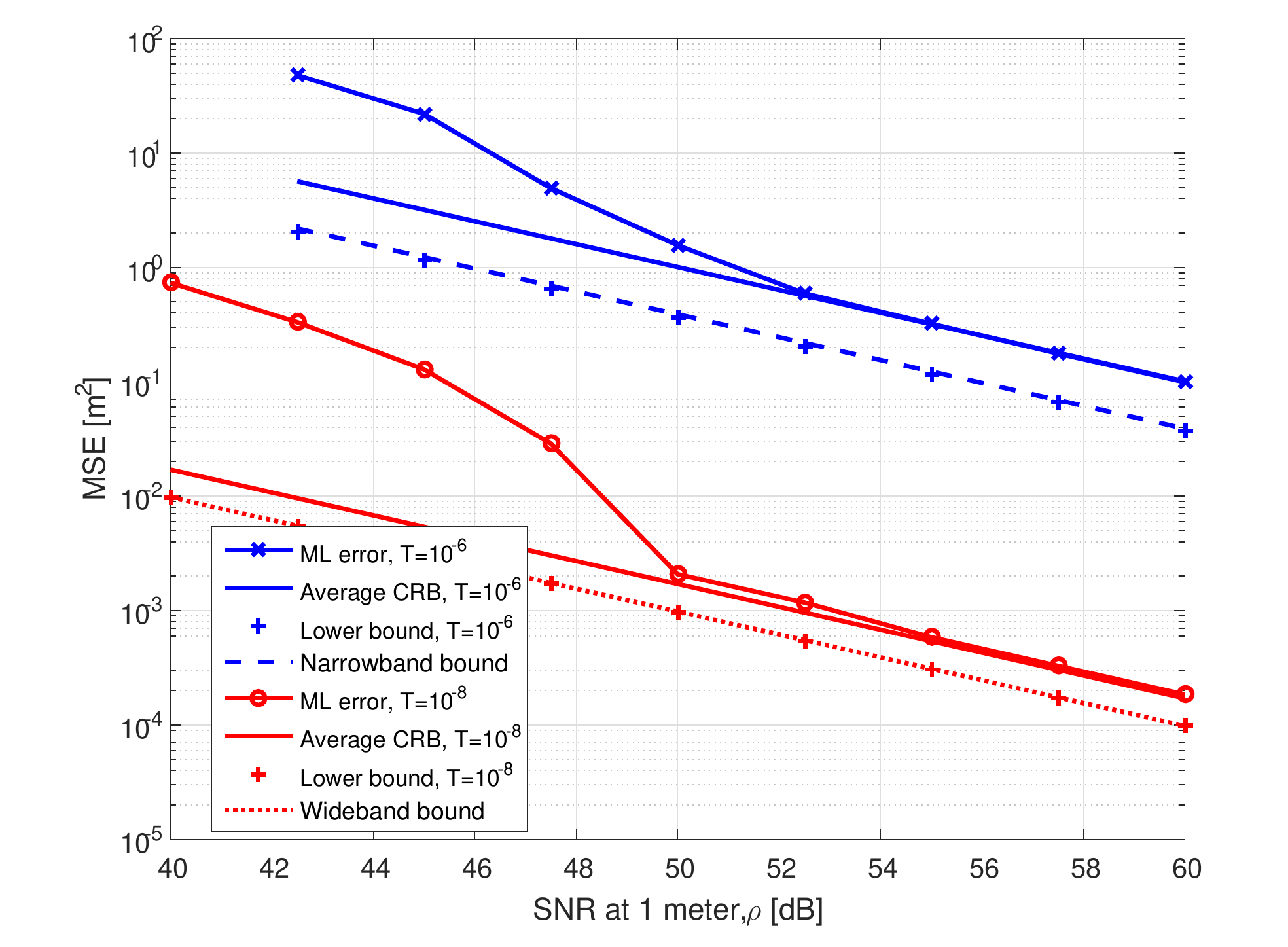}
    \caption{Source localization MSE of the ML estimator against the averaged CRB \(\ACRB(\thetavec)\)  and the lower bound \({\rm CRB}_\mathrm{LB}\) as a function of the transmitted power (measured by the SNR at 1 meter), $\lambda=0.01$m$^2$ and $\gamma=4$.}
    \label{f:vs Es}
\end{figure}
In this section we use the closed-form bounds, and show their usefulness in predicting the average localization error. All simulations  were carried out using
the pulse \begin{IEEEeqnarray}{rCl}
s(t)=\sqrt{\frac{2 E_s}{3 T k_0^2}}\cdot(1-\cos(2\pi t/T)), \ \ t\in[0,T],
\end{IEEEeqnarray}which satisfies
\(
W_{e}=\frac{4\pi^2 }{3 T^2}.\) We begin by demonstrating the behavior of the bound as a function of the SNR. Fig. \ref{f:vs Es} depicts the MSE of the maximum likelihood localizer, the average CRB, \(\ACRB(\thetavec)\)  from \eqref{ExpectedCRB},
which was approximated by averaging the
\({\rm CRB}(\thetavec,\Psimat)\) in   \eqref{e: CRB for finite} over  2000 Monte
Carlo trials with  $10^3$  sensors (instead of
infinity) deployed
randomly according to a PPP distribution at each
trial.  The figure
also depicts \({\rm CRB}_{\rm LB}\) from \eqref{e: theorem 1 bound}. The sensor-density is one  per $100$m$^2$, the path-loss exponent is $\gamma=4$ and \(T\) is
set to  $10^{-6}$ or $10^{-8}$ seconds. As expected, both bounds are inversely proportional to the SNR. The closed-form lower bound \( {\rm CRB_{\rm LB}}\) is,  indeed, not as  tight as the average CRB. Nevertheless, it exhibits exactly the same behavior, and can serve as a good indication of the actual achievable performance. The ML localizer demonstrates the well-known threshold effect, where for SNR values
  of more than $54$dB (for $1$ meter) the localizer achieves the average
 CRB bound, while for lower SNRs the localization error is significantly larger.

\begin{figure}[t]\centering
\includegraphics[width=75mm,trim={0 0 1cm 15cm},clip]{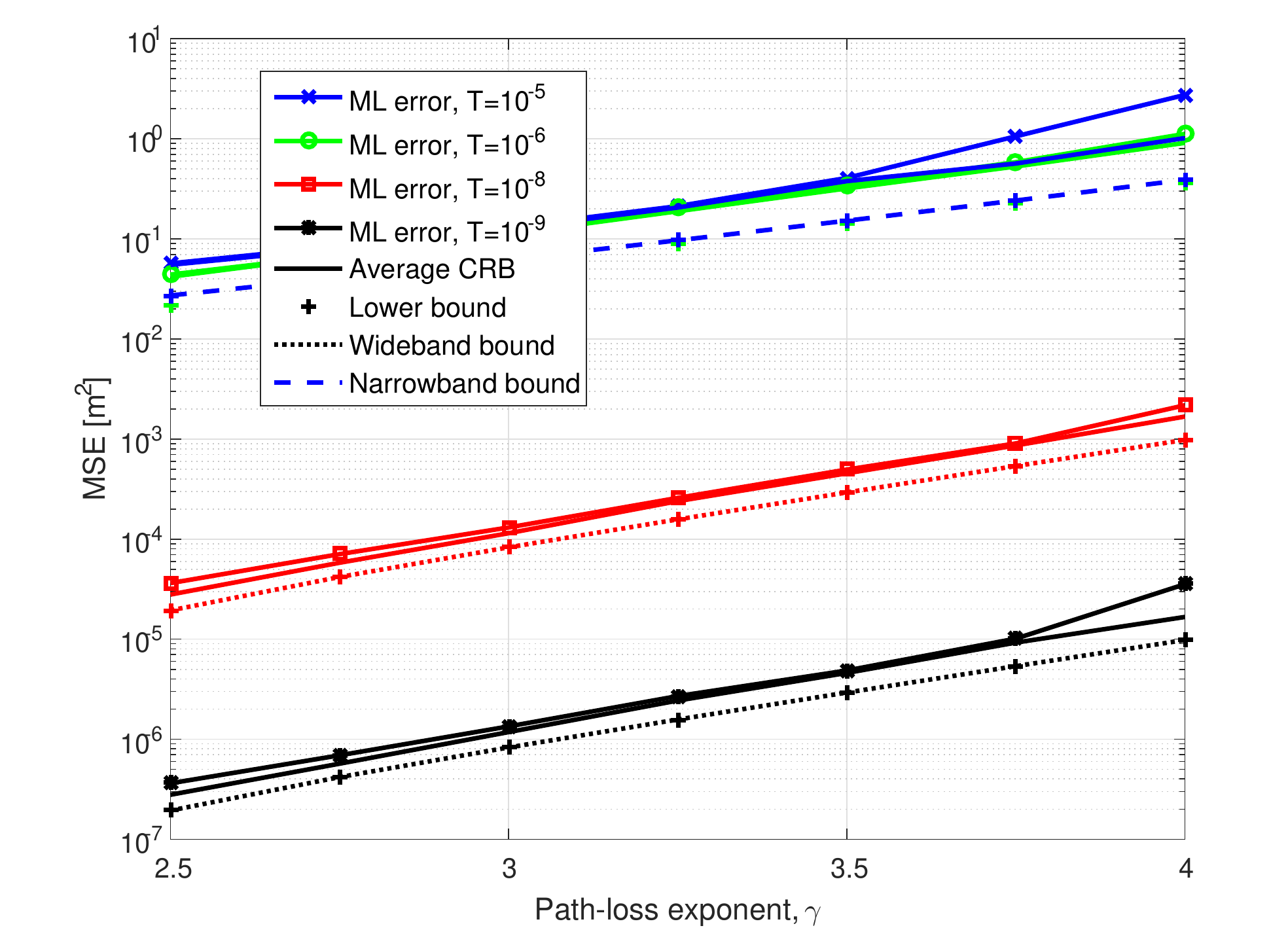}
    \caption{Source localization MSE of the ML estimator against the average CRB \(\ACRB(\thetavec)\)  and the lower bound \({\rm CRB}_\mathrm{LB}\) as a function of the path-loss exponent, $\lambda=0.01$m$^2$ and $\rho=50$dB.}
    \label{f:vs gamma}
\end{figure}
{}
 Fig. \ref{f:vs gamma} shows the localization error as a function of the path-loss exponent for the same sensor density and an SNR  of $\rho=50$ dB at one meter
away from source. Interestingly, the figure shows that for \(\gamma\leq 3.5, \) this SNR is sufficient for the ML localizer to  converge to the
average CRB, while  for higher values of $\gamma$ the  latter SNR is insufficient. As expected from \eqref{d: g func}, for small enough values, the effective bandwidth does not affect  the localization performance. In this scenario, this occurs for $T<10^{-6}$,  whereas below it, the contribution of the  time-of-arrival information becomes negligible, and the localization relies solely on the received power information. In addition, the slope of the $\log_{10}\mathrm{MSE}$ behaves differently for wideband and narrowband signals, where in the   narrowband case ($T>10^{-6}$) this slope is $d\log_{10}\mathrm{MSE}/d\gamma=0.8$ while in the  wideband case ($T<10^{-8}$), this slope is $d\log_{10}\mathrm{MSE}/d\gamma=1.2$.

\begin{figure}[t]\centering
        \includegraphics[width=75mm,trim={0 0 1cm 15cm},clip]{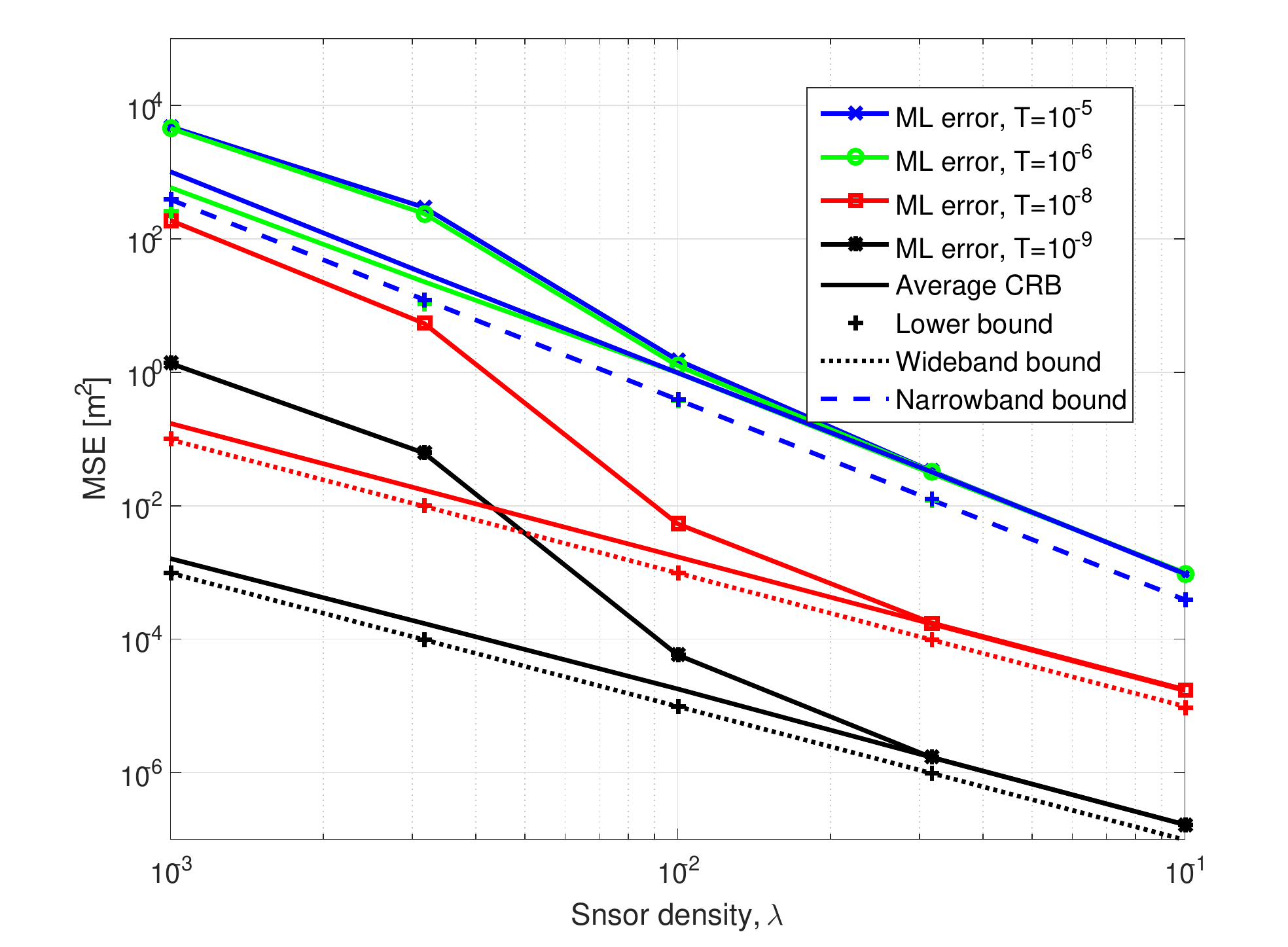}
    \caption{Source localization MSE of the ML estimator against the averaged CRB \(\ACRB(\thetavec)\)  and the lower bound \({\rm CRB}_\mathrm{LB}\) as a function of the sensor density, $\gamma=4$ and $\rho=50$dB.}
    \label{f:vs Lambda}
\end{figure}

Finally, Fig. \ref{f:vs Lambda} presents the localization performance as 
a function of the sensor density, $\lambda$, for $\gamma=4$ and $\rho=50$dB. Again, the closed-form lower bound provides
 a very good characterization of the average CRB. This figure also exhibits a difference between the slope of the MSE for narrowband and wideband signals. For narrowband signals ($T>10^{-6}$) this slope is $d\log_{10}\mathrm{MSE}/d\log_{10}\lambda=-1.5$ while for wideband signals ($T>10^{-8}$) this slope is $d\log_{10}\mathrm{MSE}/d\log_{10}\lambda=-1$. Fig. \ref{f:vs Lambda} also presents a unique threshold effect as a function of the sensor density. This threshold appears in    densities between $0.01$ and $0.03$ sensors per square meter and
 in different bandwidths. Interestingly,
this type of threshold is different than the commonly observed phenomenon, which
 occurs when  increasing the   signal  to noise ratio or   signal snapshot (which is completely equivalent to a higher signal to noise ratio) or with the number of
sensors, which increases the number of
observations. This case is different, because the number of sensors is infinite for all values of $\lambda$, but the statistics of the received signals vary with the density. Thus, at a higher density, there are typically more   sensors  close  to the target, leading to a threshold
\(\lambda\) below which the  ML estimator is characterized by large errors while
 above this threshold, it approaches
 the  average CRB.

\section{Conclusion}
\label{sec:conclusions}
We presented novel bounds on the localization MSE using  stochastic geometry. Obtaining
 these bounds required the derivation
of the  CRB in the case of an infinitely countable number of sensors, each measuring a continuous-time signal, for which we
provided a rigorous derivation, which includes an expression for the likelihood function \eqref{LikelihoodExp}. The latter likelihood can be used in the future to derive other type of bounds (see e.g.
\cite{5571899}) or to analyze similar estimation problems.   After we derived the CRB for a particular sensor locations,
by taking the expectation with respect to
a PPP distribution, we obtained the average
CRB which bounds the expected  MSE over
all possible sensor locations. This bound depends solely on the network statistics and is not a function of a particular
 sensor deployment. In addition to the average  CRB, we derived several bounds in closed-form. Both bounds provide a good characterization of the network performance, and can be used in the network planning stage to determine the performance for a given sensor density. The bounds exhibit different behavior for wideband and narrowband signals. Further research is required  to better characterize these differences. 

The  comparison of the ML estimate's MSE and the   derived bounds demonstrates
 a new and unique threshold effect, where
in the low sensor-density regime, the MSE is far above the average CRB whereas in the high sensor density regime, the MSE approaches the average CRB. This type of threshold is different from the known threshold effect with respect to signal energy and/or  the number of sensors. In this scenario, the number of sensors is infinite for all sensor densities. However, the statistics of the received signals change with the density. Thus, at a higher density, there is greater chance of having  sensors in close proximity to the target, which allows the ML error to converge to the average CRB. Further research is also required  to characterize and quantify this new threshold effect.

\appendices
{}\section{derivation of the CRB}\label{DerivationOfCRB}

 Before providing a rigorous proof for
Theorem \ref{th: CRB def}, we briefly discuss
 the 
main issues that must   be addressed
when deriving the CRB
for an infinitely
countable number of sensors.
  As stated in Section \ref{sec: CRB},
 the
CRB for  the model  \eqref{e: system model}
is already known for the case of a finite number of
sensors \citep{sieskul2009hybrid}.
   However,  extending this result to the infinite
case  is not straightforward
and requires justification and proper definitions, even if the limit does exist of the  CRB  as the number
of sensors goes to infinity. In other
words,  let \(M<\infty\) be the number of sensors located at \(\bpsi^{M}=\{(a_{m},b_{m})\}_{m=1}^M\) and \({\mathrm {CRB}}(\bth,\bps^{M})\) be the corresponding
 CRB.
    Next, let \(\psivec^{M}\) be an increasing
    sequence; i.e., \(\psivec^{1}\subset\bpsi^{2}\subset\bpsi^{3}\cdots.\) It is easy to show that \( \lim_{M\rightarrow\infty}\mathrm{
CRB}(\bth,\bps^{M})\)  is indeed a lower bound for the infinite case, where the existence of the limit
is guaranteed since \(\mathrm{CRB}(\bth,\bpsi^{M})\)
is non-negative and monotonically decreasing in \(M\). However, it is not trivial to show that the limit as \(M\) approaches
infinity is indeed the CRB bound,  \(\mathrm{CRB}(\bth,\bps)\), which  is defined with respect to the  likelihood
of the    measurements taken from the infinite number of sensors.    In other
words,
if one could calculate the CRB directly, rather than
first calculating  \(\mathrm{CRB}(\btheta,\bpsi^{M})\),
and then take the limit as \(M\longrightarrow\infty,
\)
the results would not necessarily be
equal and therefore,  \( \lim_{M\rightarrow\infty}\mathrm{
CRB}(\bth,\bps^{M})\) is not necessarily
the actual CRB.   Obtaining the actual CRB bound is beneficial, because the CRB has been studied extensively, and possesses many desired characteristics; in particular,  tightness in the high SNR regime, where it is achieved  by the ML estimator. 

    In more concrete terms,  let \(p_{\thss,\psiss^{M}}\)
and  \(p_{\thss,\psiss}\) be  the likelihoods  for estimating \(\thetavec\)  in the case of a finite and an infinite number of sensors, respectively.
 The CRB is
the inverse of the Fisher information matrix
(FIM), which is given respectively
for the finite and the infinite case 
by       
\begin{align}
\label{FiniteFIM}
\begin{array}{lll}
\Imat_{\psiss^{M}}(\bth)=\E\left\{\left.\frac{\partial
\log p_{\thss,\psiss^{M}}
}{\partial \thss^{\rm T}}\frac{\partial
\log p_{\thss,\psiss^{M}}
}{\partial \thss}\right\vert \bps^{M} \right\}\end{array}\\
 \label{InfiniteFIM}
\begin{array}{lll}\Imat_{\psiss}(\bth)=\E\left\{\left.
{\frac{\partial
\log p_{\thss,\psiss}}{\partial \thss^{\rm T}}\frac{\partial
\log p_{\thss,\psiss}}{\partial \thss}
}\right\vert\bps\right\}\end{array};
\end{align}
 By definition, to derive  the CRB for the infinite
case, one has to calculate \(\Imat_{\psiss}(\bth)\)
directly according to \eqref{InfiniteFIM};
i.e., to define the infinite likelihood,
\(p_{\thss,\psiss}\), and substitute
it in \eqref{InfiniteFIM}. Another alternative would be to calculate \eqref{InfiniteFIM} by taking the limit of  \eqref{FiniteFIM}      as $M\longrightarrow\infty$;  i.e.,
  \beq \label{convergenceOfMarginaFIM}\lim_{M\to\infty}\Imat_{\psiss^{M}}(\bth)=\Imat_{\psiss}(\bth).
 \eeq  However, for \eqref{convergenceOfMarginaFIM} to be true,   the likelihood  \(p_{\thss,\psiss^{M}}\)  must converge to \(p_{\thss,\psiss}\) as \(M\longrightarrow\infty\), and similarly   for the derivative of the log-likelihood with respect to \(\btheta,\) as well as the product of the derivatives.  If
these conditions are satisfied, it is possible to calculate the CRB via \eqref{convergenceOfMarginaFIM}.

This proof has several parts;  some of which  are  cumbersome. Hence, to make  it  tractable, we begin with   a high level description of each part. In Part \ref{PartA}, we define the probability space on which the observations   from a finite number of sensors are defined as random elements. We then define the likelihood in this case. In Part   \ref{PartB}, we extend the probability space to the case of an infinite number of sensors; i.e., we construct a probability space such that the observations from all the sensors constitute a sample point. This is  necessary  to obtain a well defined  likelihood function, which is used to estimate \(\thetavec\). In Part   \ref{PartC}, we derive, in closed-form,  the likelihood in the case of an infinite number of sensors \(p_{\thss,\psiss}\), which is then used  in  Part  \ref{PartD} (Lemma \ref{Lemma7}) to derive   a formula for the  FIM in the case of an infinite number of sensors with additive white Gaussian noise. It turns out, from Lemma \ref{Lemma7}, that the infinite FIM can be calculated via  \eqref{convergenceOfMarginaFIM}.  In Part \ref{PartE} we substitute the model, \eqref{e: system model}, into the formula for the FIM and calculate the CRB.    
\begin{IEEEproof}[Proof of Theorem \ref{th: CRB def}]\\
\part{Definition of the probability space and the likelihood function   in the case of a finite number of sensors:}
\label{PartA}
% \textcolor{blue}{Without loss of generality
% we assume
%  that \(D_{m}\leq D_{m'}\) for every
%  \(m\leq m'\).
% }We consider the observations \(\{\{r_{
% \psi_{ m}}(t): t\in[0,T]\}\}_{ m=1}^{\infty}\)
% as a random point 
% in   \({ \real }^{ [0,T] \times\psi}\);
% i.e., each   point is a countable
% set of  functions from \([0,T]\) to \(\real\). 
We start with  some definitions
and known results, which
will later be used in the derivation. We begin with defining a suitable probability space in which the likelihoods are defined.   Because white Gaussian noise does not exist  as an ordinary mathematical stochastic process,  we consider the integral of \eqref{e: system model} and replace the model
of \eqref{e: system model} with the 
 equivalent model   (see e.g. \citep{poor1994introduction},
 Sections VI.C.2 and VII.B)\beq 
 \label{AlternativeSignalModel} R_{  m}(t)=
k_0D_m^{-\gamma /2}\int _{0}^{t}s\left(u-\tau _m \right)du+W_m(t),~~ t\in \real_{+}, \eeq  where   \(W_{ m}(t)\) is a Wiener 
 process 
%  \footnote{ The signal model
% in \eqref{AlternativeSignalModel}
%     \[ dR_{ \psi_{ m}}(t)=
% k_0D_m^{-\gamma /2}s\left(t-\tau _m
%  \right)dt+dW_m(t)\]  }
with covariance \( \E(W_{ m}(t)W_{ m}(u))= N_{ 0}/2 \min\{t,u\}\). Because the Wiener process is continuous a.s.,   for each \(m,\)   \(W_{ {m}}(\cdot)\) can be seen as an element in \({\mathbb S}\), the space of continuous functions on \(\real_{}\); i.e.,  \({\mathbb S}={\cal C}(\real_{})\). Furthermore, because \(s(t-\tau_{})\) is
integrable for every \(\tau\),  each \(R_{m}
(\cdot)\) is also an element in ${\mathbb S} $. The space
\({\mathbb S}\) is a complete separable metric space, dubbed  Polish space (see e.g. \citep{shreve1991brownian},
Sec. 2.4),  with respect to the  distance \(d(f_{ 1},f_{ 2})=\sum_{n=1}^{\infty}\frac{1}{2^{n}}\sup_{ t\in[0,n]}\vert f_{ 1}(t)-f_{ 2}(t)\vert\).  Next, consider the measurable  space  \(({\mathbb S},{\cal B}({\mathbb S}),\mu)\), where \({\cal B}({\mathbb S})\) is the Borel \(\sigma-\)algebra and  \(\mu \) is the Wiener measure. Given a node location, $\psi_m$, each observation \(R_{ m}(\cdot)\in{\mathbb S}\) is a random element in \(({\mathbb S},{\cal B}({\mathbb S}),P_{ \thss,{\psi_{ m}}})\), where \(P_{ \thss,{\psi_{ m}}}\) is the law of \(R_{ m}(t)\)\footnote{In our case  \(P_{ \thss,{\psi_{ m}}}(R_{ m}(t)\in B)= \mu(B-s(t-\tau_m))\) where \(B-s (t-\tau_m)=\{x(t)\in {\mathbb S}: x(t)-s(t-\tau_m)\in B \}.\)}. Now that the probability space has been defined,  to  define a likelihood, one  needs a measure with respect to which  \(P_{ \thss,\psi_{ m}}\) is absolutely continuous. Such a measure is \(\mu\)   if \(\int_{ 0}^{\infty}s^2(t-\tau_m)dt<\infty,\)  and  the  likelihood function for estimating \(\thetavec\)  from the observation \(R_{ m}\),   given \(\psi_{ m}\) can be written as   \citep{kailath1969general}
\beqI
\label{likelihoodFunction}p_{ \thss,\psi_{m}}
(R_{ m}(t))&=& \frac{dP_{ \thss, \psi_{ m}}}{d\mu}(R_{ m}(t))
 \TwoOneColumnAlternate{\\ \nonumber
&&\hspace{-21mm}}{}= \exp\left\{\frac{2}{N_{ 0}} \int_{ 0}^\infty s_{ m}(t;\thetavec)d R_{ _{ m}}(t)- \frac{1}{N_{ 0}}\int _{ 0}^{\infty}s_{ m}^2(t;\thetavec) dt\right\}
\eeqI
 where \({d P_{ \thss,\psi_{ m}}}/{d\mu}\) denotes the Radon-Nikodym derivative, and \beq \label{Defines_m}s_{m}(t;\bth)=k_0D_m^{-\gamma /2}s\left(t-\tau _m \right).\eeq Furthermore, consider the observation \(\Rmat_{\Mcal} \define\{R_{  m}(t)\}_{ m\in\Mcal }\in {\mathbb S^{|\Mcal |}}\) where $\Mcal=\{\alpha_{1},\alpha_{2},\cdots,\alpha_{M}\} \subset \nat$. Because  \(W_{m}(t), W_{n}(t)\) are independent processes for every \(m\neq n\),  \(\Rmat_{\Mcal }\) is naturally represented as a random element in \((\mathbb S^{|\Mcal |},{\cal B}({\mathbb S}^{|\Mcal |}),P_{\thss,\psiss_\Mcal })\), where \(P_{\thss,\psiss_\Mcal }\) is the
 product measure    of \(\{P_{\thss,\psi_{m}}\}_{m\in{\cal M}}\),  with  the product defined  in the usual way (see e.g.   \citep{Folland}). The likelihood of  the \(\Rmat_{\cal M}\) given \(\bpsi_{\Mcal }=\{\psi_{m}:m\in\Mcal \}\)  is    
\begin{IEEEeqnarray}{rCl}
\label{LikelhihoodM}
\begin{array}{lll}
p_{ \thss,\psiss_{\Mcal}}(\Rmat_\Mcal )
=\exp\left\{ \frac{2}{N_{ 0}}\sum_{ m\in\Mcal }{\displaystyle\int}_{ 0}^\infty s_{ m}(t;\thetavec)d R_{ _{ m}}(t)\cond{\right.\\[1em]~~~~~~~~~~~~~~~~~~~ \left.}
- \frac{1}{N_{ 0}}\sum_{ m\in\Mcal } {\displaystyle\int} _{ 0}^{\infty}s_{ m}^2(t; \thetavec) dt\right\} 
\\[1em]~~~~~~~~~~~~~~~
=\prod_{m\in M}
p_{ \thss,\psi_{m}}
(R_{ m}(t)).
\end{array}
\end{IEEEeqnarray}

\part{The probability space and likelihood function in the case of an infinite number   of sensors:} \label{PartB} Let $ \Rmat_{}\define\{R_{m}(t)\}_{m=1}^{\infty}$; to derive the likelihood for an infinite
number of  sensors we need  a probability space \((\Omega_{\psiss},{\cal F_\psiss},P_{\thss,\psiss})\) on which  \(\Rmat_{}\) is a sample point. 
This can be accomplished by using  Kolmogorov's consistency theorem (see
e.g., \citep{athreya2006measure},
Theorem
6.3.1 and Remark 6.3.4), but first  we need  the following definition.\begin{definition}
Let \(\Mcal=\{\alpha_{1},\alpha_{2}\cdots\alpha_{M}\}\subset\nat\),   \(M<\infty\), with \(\alpha_{i}\neq \alpha_{j},~\forall i\neq j\). A finite dimensional cylinder is a set   $C{_\Mcal}\subset \sd^{\infty}$ where there exists \(B\in\Bcal(\sd^{M})\) such that  every \((x_{1},x_{2},x_{3}\cdots)\in C{_\Mcal}\) satisfies $\xvec_{\Mcal}\define(x_{\alpha_{1}},x_{\alpha_{2}}\cdots x_{\alpha_{M}})\in B$.  The set  \(B\) is called the basis of the cylinder \(C_\Mcal\).\end{definition}
   
   Next, from Kolmogorov's consistency theorem  it follows that for every \(\bpsi\) there exists  a probability space  \((\Omega_{\psiss},{\cal F_{\psiss}},P_{\thss,\psiss})\), where \(\Omega_{\psiss}=\sd^{\infty}\),
 $\Fcal_{\psiss}$ is the sigma algebra generated by the set of finite dimensional  cylindrical sets and   for every  finite dimensional  cylinder \({ C}\in\Fcal_{\psi}\) with basis  \(B \)  the following is satisfied   \beq\label{Margian} P_{\thss,\psiss}(C _\Mcal )=P_{\thss,\psiss_{\Mcal} }( B). \eeq   Similarly, there exists a  measure \(P_{0}\) on \((\Omega_{\psiss},\Fcal_{\psiss})\)  such that \begin{equation}\label{DefineP_0}P_{0}(R_{\psiss_{\Mcal} }\in B)=\mu^{|\Mcal |}(B).\end{equation}
Because \(P_{\thss,\psiss_\Mcal }\ll\mu^{\vert \Mcal\vert}\)
for every \(\Mcal \subset \nat\) with \(|\Mcal |<\infty\), and since both \(P_{0}\) and \(P_{\thss,\psi}\) are defined using the Kolmogorov consistency theorem, it follows that \(P_{\thss,\psiss}\ll P_{0}\).\footnote{This is because   every \(C\in\bcal(\sd^{\infty})\) with $P_{0}(C)=0$
can be written as $C=\bigcup_{n=1}^{\infty}C_{n}$
where $C_{n}$ are cylindrical sets with
bases \(B_{n}\in\bcal(\sd^{k_{n}}),\)
$k_{n}<\infty,~\forall n$. Now, by subadditivity,
\(P_{0}(C_{n})=0\), thus  \(\mu^{k_{n}}(B_{n})=0
\Rightarrow
P_{\thss,\psiss_{A_{n}}}(B_{n})=0\)
for every \(A_{n}\subset\nat, \vert
A_{n}\vert=k_{n}\), 
and therefore, by subadditivity \(P_{\thss,\psiss}(C)=0.\)  }  

Now that we have a well defined probability
space \((\Omega_{\psi},{\cal F_{\psi}},P_{\thss,\psiss})\),
and a measure \(P_{0}\) which dominates \(P_{\thss,\psiss}\),  the likelihood  of \(\Rmat_{}\)      \beq
   \label{LikelihoodPsi}
p_{\thss,\psiss}(\Rmat_{})=\frac{dP_{\thss,\psiss}}{d  P_{0}}(\Rmat_{}),
\eeq  
    is well defined and  the CRB is given by \(CRB(\bpsi)=\Imat^{-1}(\thetavec,\bpsi),\) where, \beq\label{FisherByDefinition}  \Imat(\thetavec,\bpsi)=\E\left\{  \frac{\partial\log p_{\thss,\psi}(\Rmat_{})}{\partial\thetavec^{\rm T}} \frac{\partial\log p_{\thss,\psi}(\Rmat_{})}{\partial\thetavec}\right\}.
\eeq  

\part{Derivation of the likelihood \(p_{\thss,\psi}\) in closed-form:} \label{PartC} We now derive the likelihood \(p_{\thss,\psiss}(\Rmat_{})\)
 in closed-form by showing that the marginal likelihood \(p_{\thss,\psiss_{\Mcal }}\)  converges w.p. 1 to \(p_{\thss,\psiss}\). If the observations \(\{R_{m}(t)\}_{m=1}^{\infty}\) were  random variables (i.e., for each
\(m\), \(R_{m}\) was a  Borel measurable function from \(\Omega\) to \(\real\)),  the
convergence of the marginal  likelihood
to the infinite one  would follow immediately from Grenander's theorem (See \citep{grenander1981abstract}, Chapter 3, Corollary 1). In our problem,
however,  \(R_{m}(\cdot)\) is a random element on \((\sd,\bcal(\sd),\mu)\); thus, it is necessary to  extend  Granander's theorem to the case at hand.  \begin{theorem}\label{th: likelihood limit} Let \(\{{\cal M}_{\nu}\}_{\nu=1}^\infty\subset\nat\)
be an increasing sequence of sets (that
is  \({\cal M}_{1}\subset {\cal M}_{2}\subset\cdots\)),
and for each \(\nu\), define  \(
p_{\thss,\psiss_{\Mcal_{\nu}}}\) as in     
  \eqref{LikelhihoodM}. Consider \(p_{\thss,\psiss}\)    
defined in \eqref{LikelihoodPsi}, then,   \beq\label{ConvergensOfMarginalLikelihood}
p_{\thss,\psiss}(\Rmat_{})=\lim_{\nu \to\infty}
p_{\thss,\psiss_{\Mcal_{ \nu}} }(\Rmat_{{\Mcal }_{\nu}}),~ {\rm
w.p.~} 1.
\eeq  
 \end{theorem}
 
\begin{IEEEproof}   For each \({\cal
M}_{\nu}\), let    \(C_{\nu}\subset\sd^{\infty}\)   be a finite dimensional cylinder with basis \((\Mcal_{\nu},B_{\nu}),\) where  \(B_{\nu}\in \Bcal(\sd^{|\Mcal_{\nu}|}  )\). Then,  by the Radon-Nykodym theorem 
\begin{IEEEeqnarray}{rCl}
P_{\thss,\psiss}(C_{\nu})&=&\int_{C_{\nu}}p_{\thss, \psiss}(\xvec)dP_{0}(\xvec)\TwoOneColumnAlternate{\nonumber \\ &=&}{=}\int_{B_{\nu}} p_{\thss,\psiss_{\Mcal_{\nu} }}(\xvec_{{\Mcal_{\nu} }})d\mu^{|\Mcal_{\nu}|}(\xvec_{\Mcal_{\nu} }). 
\end{IEEEeqnarray}

Since this holds for every \(B_{\nu}\in\Bcal(\sd^{\vert \Mcal_{\nu}\vert}),\) it follows that  \beq p_{\thss,\psiss _{\Mcal_{\nu} }}(\xvec)=\E_{P_{0}}\{p_{\thss, \psiss}(\xvec)\vert \fcal_{\Mcal_{\nu} }\},\eeq where  
 \(\fcal_{\Mcal_{\nu} }=\sigma({\cal C}_{\cal N}),\) where \({\cal C}_{\cal N}\) is
the collection of all finite dimensional
cylindrical sets, with basis \(({\cal
N},B)\), such that \({\cal N}\subset \nat\) and \(B\in{\cal B(\sd^{\vert
{\cal N}\vert})}\).  Note that
 \(\{\fcal_{\Mcal_{i}}\}_{i=1}^{\infty}\) is a filtration since \(\Mcal_{1}\subseteq\Mcal_{2}\subseteq\cdots\). Thus, \(p^{}_{\thss,\psiss_{\Mcal}}\) is an abstract Doob Martingale (see \citep{chatterji1964note}, Theorem
4) for every \(\Mcal_{1}\subseteq\Mcal_{2}\subseteq\cdots\), and therefore converges to
\(p_{\thss,\psi}\) a.s.-\(P_{0}\), which establishes the desired result (recall that \(P_{0}\) dominates \(P_{\thss,\psi}\)).
\end{IEEEproof}
To summarize, Theorem \ref{th: likelihood limit} provides a close-form expression for the likelihood 
\beqI
\label{LikelihoodExp}
p_{\thss,\psi}(\Rmat_{})=e^{\frac{1}{N_{0}} {\displaystyle\sum\limits_{m=1}^{\infty}} \left(2{\displaystyle\int}_{0}^\infty s_{m}(t; \thss)d R_{_{m}}(t)-{\displaystyle\int} _{0}^{\infty}s_{m}^2(t;\thss) dt\right)}
\eeqI  
\part{Formula for the Fisher Information matrix   in the case of  an infinite number of sensors:} \label{PartD} Now that we have the likelihood in  closed-form, 
we can derive the FIM, and show that (\ref{convergenceOfMarginaFIM}) is satisfied.
 To this end we need the following lemma,
which extends the well known formula for the
 FIM  in the case where the observation is a finite set  of continuous signals, each with  additive Gaussian
 white noise, to the case of  infinitely countable continuous signals.\begin{lemma}
\label{Lemma7} If  \(s(t)\) is bounded, differentiable, with a bounded derivative, then  for every \(\bpsi\) which satisfies \eqref{FiniteSum}, the \(i,j\) entry of the \(2\times2\) FIM is given by \beq\label{SumOfMarginalFIM}[\Imat(\thetavec)]_{i,j}= \frac{2}{N_{0}}\sum _{m=1}^{\infty } \int _{0}^\infty\frac{\partial s_m(t;\btheta )}{\partial \theta _i}\frac{\partial s_m(t;\btheta )}{\partial \theta _j}dt\eeq
\end{lemma}

\IEEEproof To prove the lemma, one must prove that \eqref{FisherByDefinition}
reduces to \eqref{SumOfMarginalFIM}.
In order to evaluate      \eqref{FisherByDefinition},
we need the following proposition: 
\begin{proposition} \label{DerivativeOfTheLikelihood}   Consider the
likelihood \(p_{\thss,\psiss}(\Rmat)\)
in \eqref{LikelihoodExp},
then  \(\log p_{\thss,\psiss}(\Rmat)\in
 L_{2}(\Omega_{\psiss},\fcal_{\psiss},P_{\psiss})
 \) and 
 \beq\label{derivativeUnderInfiniteSum}\frac{\partial \log p_{\thss, \psiss}}{ \partial\theta_{j}} =\sum_{m=1}^{\infty}\frac{\partial\log p_{\thss,\psi_{m}}(R_{m}(t)) }{\partial\th_{j}}\in L_{2} (\Omega_{\psiss},\fcal_{\psiss},P_{\psiss})\eeq
\end{proposition}
where the equality is in the mean-square sense; i.e., with respect to the norm of \(L_{2}(\Omega_{\psi},\fcal_{\psi},P_{\psi})\). \IEEEproof From \eqref{LikelihoodExp},   
\beqI
\label{44}
{\log p_{\thss,\psiss}(\Rmat) }= \sum_{m=1}^{\infty} \log( p_{\thss,\psi_{m}}(R_{m}(t))),
\eeqI 
and from \eqref{likelihoodFunction}  
\beqI\label{e: log p phim}
\log( p_{\thss,\psi_{m}}(R_{m}(t)))&=&\frac{2}{N_{ 0}} \int_{ 0}^\infty k_{0}D^{-\gamma/2}_{m}s(t-\tau_{m}(
\thetavec
))d R_{ _{ m}}(t)\Dcond{\nonumber \\&&}{}- \frac{1}{N_{ 0}}\int _{ 0}^{ \infty
}k_{0}D_{
m}^{-\gamma}s_{ }^2(t-\tau_{m}(\bth)) dt\eeqI 
Because \(s(t)\) is bounded and  square integrable, it follows that 
\beq\label{defineZm} Z_{m}\define\int
_{ 0}^\infty s(t-\tau_{m}(\thetavec))d R_{ _{ m}}(t)\in L_{2},\eeq and  \beq \label{definexim}\xi_{m}\define\int_{0}^ {\infty}s^{2}(t-\tau_{m}(\btheta))<\infty. \eeq Thus, \(\log(p_{\thss, \psi_{m}}(R_{m}))\in L_{2}\). Furthermore, because \(\int _{0}^{\infty}s^{2}(t)dt<\infty\), there exists a constant, $u$, (which is not a function of \(\bth\) and $m$)  such that \(\E\{Z_{m}^{2}\}\leq u\),  and since     
 \(\sum_{m=1}^{\infty}D_{m}^{-\gamma}<\infty\), it follows that \beq\label{boundedness}  \log p_{\thss,\psi}(\Rmat_{\psi})= \sum_{m=1}^{\infty}\frac{1}{ N_{0} }(2 k_{0} D_{m}^{-\gamma/2}Z_{m}+D_{m}^{-\gamma}\xi_{m})\in L_{2} \eeq This establishes the first part of the
proposition. For the second part, we note \(s'(t)\define ds(t)/dt\) exists  bounded,   square integrable with
a bounded derivative, and hence also \({\partial s_{m}(t;\bth)}/{\partial\bth _{j}}\). Thus,  the order of the  derivative   and the integration  in (\ref{e: log p phim}) can be exchanged\footnote{This is possible due to the dominant convergence theorem for stochastic integrals (see e.g., \citep{prigent2003weak} Theorem 1.1.19), and the regularity conditions, \eqref{eq:RegularityConditionSquareIntegrableFirstAndSecondDerivative}.
It is then possible to justify it similar to \citep{Folland},
theorem 2.27.}:
\beq\begin{array}{ll}
 \frac{\partial\log p_{{\thss},\psi_{m}}(R_{m}) }{\partial\th_{j}}\Dcond{\\[1em]~~~}{}=\frac{1}{N_{0}}  \left(2\intarr_{0}^\infty \frac{\partial s_{m}(t; \thss)}{\partial \th_{j}}d R_{_{m}}(t)-\intarr _{0}^{\infty}\frac{\partial s_{m}^2(t;\thss)}{\partial \theta_{j}} dt\right),
\end{array}\eeq
which can be written as (see e.g., \citep{poor1994introduction} sec VII.B)
\beq\label{SimplifyDerivativeMarginal}
\frac{\partial\log p_{\thss,\psi_{m}}(R_{m}) }{\partial\th_{j}}=\frac{2}{N_{0}} \left(\int_{0}^\infty \frac{\partial s_{m}(t; \thetavec)}{\partial \theta_{j}}d W_{_{m}}(t)\right).
\eeq
Note that \(\vert\partial D_{m}/\partial \th_{j}\vert\leq 1\); thus, \(\vert\partial D_m^{-\gamma /2}/\partial \th_{j}\vert=\vert D_m^{-\gamma /2-1} \partial D_{m}/\partial \th_{j}\vert\leq D_{m}^{-\gamma/2-1}\).  Denote \(\partial D_{m}/\partial \th_{j}=D_{m,j}'\), then  
\begin{align}\nonumber
\frac{\partial\log p_{\thss,\psi_{m}}(R_{m}) }{\partial\th_{j}}= &\frac{2}{N_{0}} \left(\int_{0}^{\infty} [D_m^{-\gamma /2-1}\frac{\partial D_{m}}{\partial \theta_{j}}s(t -\tau_{m}(\bth))\right.  \Dcond{\\\nonumber&}{}\left.+\frac{D_m^{-\gamma /2}}{c}\frac{\partial D_{m}}{\partial \theta_{j}}s'(t-\tau_{m}(\bth))] d W_{_{m}}(t)\right) \\\label{derivativeODm}=&(O( D_{m }^{-\gamma/2})(S_{m}(\bth)+S'_{m}(\bth)),
\end{align} 
where
\begin{align}
S_{m}(\bth)\define&\left(\int_{0}^\infty s(t-\tau(\bth))d W_{_{m}}(t)\right)\in L_{2} \\S'_{m} (\bth)\define&\left(\int_{0}^\infty s'(t-\tau(\bth))d W_{_{m}}(t)\right)\in L_{2}
.\end{align} Hence 
\beq
\frac{\partial\log p_{\thss,\psi_{m}}(R_{m})}{\partial\th_{j}}\in L_{2}
\eeq
Recalling that
  \(\sum_{m=1}^{\infty}D_{m}^{-\gamma}<\infty\) and using  (\ref{44}),
it follows that 
\beq\label{infiniteSumOfLogDerivatives}
\sum_{m=1}^{\infty}\frac{\partial\log p_{\thss,\psi_{m}}(R_{m}(t)) }{\partial\th_{j}}\in L_{2} (\Omega_{\psiss},\fcal_{\psiss},P_{\psiss})
\eeq 
 It remains to show that \(p_{\thss,\psi}(\Rmat)\) is differentiable and that its derivative is given by \eqref{infiniteSumOfLogDerivatives}. Using \eqref{boundedness} and  \(\sum_{m=1}^{\infty}D_{m}^{-\gamma}<\infty\), \eqref{44} can be written as  \beq
\log p_{\thss,\psi}(\Rmat_{})=\sum_{m=1}
^{M}\log(p_{\thss,\psi_{m}}(R_{m}))+\tilde U_{M}(\btheta),\eeq where \beq \tilde U_{M}= \sum_{m=M+1}^{\infty}\frac{k_{0}D_{m}^{-\gamma/2}}{ N_{0} }(2Z_{m}+\xi_{m})\in L_{2} .\eeq 
% \beq\begin{array}{ll}
%  \frac{\log p_{\thss+\Delta\th_{j},\psi}(\Rmat_{\psi})-\log p_{\thss,\psi}(\Rmat_{\psi})}{\Delta \theta_{j}}-\sum_{m=1}
% ^{M}\frac{\partial\log(p_{\thss,\psi_{m}}(R_{m}))
% }{\partial\th_{i}}-
% \end{array}\eeq

Thus,  
\beq\label{53} \frac{\Delta_{j}(\log p_{\thss,\psi}(\Rmat_{}))}{\Delta \theta_{j}}
=\sum_{m=1}
^{M}\frac{\Delta_{j}(\log(p_{\thss,\psi_{m}}(R_{m})))}{\Delta \theta_{j}}+ \frac{\Delta_{j}( \tilde U_{M}(\btheta))
}{\Delta \theta_{j}}\eeq
  where \(\Delta_{j}\) is an operator defined on functions from \(\real^{2}\) to \(\real\), such that for a given \(g:\real^{2}\rightarrow \real\),      \(\Delta_{1}(g(\theta_{1},\theta_{2}))\define g(\th_{1}+ \Delta\theta_{1},\theta_{2})\) and \(\Delta_{2}(g(\theta_{1}, \theta_{2}))\define g(\th_{1},\theta_{2}+ \Delta\theta_{2})\). 
  We now show that \({\Delta_{j}( \tilde U_{M}(\btheta))
}/{\Delta \theta_{j}}\) converges to zero as \(M\to\infty\) and \(\Delta\theta_{j}\rightarrow 0\).  From \eqref{defineZm}, \eqref{definexim} and \(\sum_{m=1}^{\infty}D_{m}^{-\gamma}<\infty, \) it follows that for every \(\epsilon>0\) there exists \(M_{0}(\epsilon)\), such that \(\forall M\geq M_{0}(\epsilon)\), $\E\{\tilde U_{M}^{2}(\bth)\}<\epsilon$ for every \(\btheta\). Thus, by letting \(\epsilon=o(\Delta\theta_{j})^{2}\) we have    
 \begin{align}\label{69eq}
  \frac{\Delta_{j}(\log p_{\thss\psiss}(\Rmat_{}))}{\Delta \theta_{j}}&=\sum_{m=1}
^{M_{0}}\frac{\Delta_{j}(\log(p_{\thss,\psi_{m}}(R_{m})))}{\Delta \theta_{j}}+o_{{\rm ms},\Delta \theta_{j}}(1),\end{align} where \(o_{{\rm ms},\Delta \theta_{j}}(1)\) denotes a random function \(Q(\Delta\theta_{i})\) which converges to zero in the mean square sense as \(\Delta \theta_{j}\longrightarrow 0\). Denote  
\begin{IEEEeqnarray}{lll}
W_{M_{1},\Delta\theta_j}\Dcond{\\\nonumber}{}\define\sum_{m=1}
^{M_1}\frac{\Delta_{j}(\log(p_{\thss,\psi_{m}}(R_{m})))}{\Delta \theta_{j}}
-\sum_{m=1}
^{M_{1}}\frac{\partial\log(p_{\thss,\psi_{m}}(R_{m}))
}{\partial\th_{j}}.
\end{IEEEeqnarray}
Then, substituting into \eqref{69eq}, we obtain
\begin{align}
\begin{array}{lll}
  \frac{\Delta_{j}(\log p_{\thss,\psiss}(\Rmat_{}))}{\Delta \theta_{j}}=\sum\limits_{m=1}
^{M_{1}}\frac{\partial\log(p_{\thss,\psi_{m}}(R_{m}))
}{\partial\th_{j}}\Dcond{\\[1em]~~~~~~~~~~~~~~~~~~~~}{}+W^{}_{M_{1},\Delta\theta_{j}}+o_{{\rm ms},\Delta\theta_{j}}(1),
\end{array}\end{align} 
Next, by the mean value theorem, for every \(\Delta\theta_{j}\) there exists \(\theta_{j,m}\) between  \(\theta_{j}\) and \(\theta_{j}+\Delta\theta_{j}\) such that  
\beq\label{MeanValueTheorem}\begin{array}{ll}
V_{M_{1},\Delta\theta_{j}}\define\sum\limits_{m=M_{1}}
^{M_{0}}\frac{\Delta_{j}(\log(p_{\thss,\psi_{m}}(R_{m})))}{\Delta \theta_{j}}\Dcond{\\[1em]~~~~~~~~~}{}=\sum\limits_{m=M_{1}}
^{M_{0}}\left.\frac{\partial\log(p_{\thss,\psi_{m}}(R_{m}))
}{\partial\th_{j}}\right\vert_{\theta_{j}=\theta_{j,m}}
\end{array}\eeq
Thus,\beq\label{EQ:LogLikelihoodExpression}
\begin{array}{ll} \frac{\Delta_{j}( \log p_{\thss,\psiss}( \Rmat_{}))}{\Delta \theta_{j}}=\sum\limits_{m=1}
^{M_{1}}\frac{\partial\log(p_{\thss,\psi_{m}}(R_{m}))
}{\partial\th_{j}}\Dcond{\\[1em]~~~~~~~~~~~~~~~~~~~~}{}+W_{M_{1},\Delta_{\theta_{j}}}+V^{}_{M_{1},\Delta_{\theta_{j}}}
+o_{{\rm ms},\Delta\theta_{j}}(1),\end{array}
\eeq
from which it follows that  
\beq
\begin{array}{ll} \left\Vert\frac{\Delta_{j}( \log p_{\thss,\psiss}( \Rmat_{}))}{\Delta \theta_{j}}-\sum\limits_{m=1}
^{M_{1}}\frac{\partial\log(p_{\thss,\psi_{m}}(R_{m}))
}{\partial\th_{j}}\right\Vert_{\rm m.s.}\Dcond{\\[1em]~~}{}\leq \Vert W_{M_{1},\Delta\theta_{j}}\Vert_{\rm ms}+\Vert V^{}_{M_{1},\Delta{\theta_{j}}}
\Vert_{\rm m.s.}+o_{{\rm m.s.},\Delta\theta_{j}}(1)\end{array}
\eeq
 
 Examining the last equality, we observe  that  \(\lim_{M_{1}\rightarrow\infty}\lim_{\Delta\theta_{j}\rightarrow\infty}W_{M_{1},\Delta_{\theta_{j}}}=0\) in the m.s. As for   \(V_{M_{1},\Delta \theta_{j}}\),  using \eqref{derivativeODm}, it can be written as  \beq\label{MeanValueTheorem}\begin{array}{ll}
V_{M_{1},\Delta\theta_{j}}=\sum\limits_{m=M_{1}}
^{M_{0}}\left((S_{m} ( \thetavec) +S'_ {m}(\thetavec))\vert_{\theta_{j}=\theta_{j,m}}\right) \Dcond{\\[1em]~~~~~~~~~~~~~~~~~~~~~~~~~~~~\times}{}O((D_{m}(\thetavec))^{-\gamma/2}\vert_{\theta_{j}=\theta_{j,m}}).
\end{array}\eeq
Because \(S_{m}(\thetavec)+S'_{m}(\thetavec)\in L_{2}\) for every \(\thetavec\), in order to show that \(\lim_{M_{1}\rightarrow\infty}\lim_{\Delta\theta_{j}\rightarrow\infty} V_{M_{1},\Delta\theta_{j}}=0\)  in the m.s., it is sufficient to show that \(\sum_{m=1}^{\infty} (D_{m}(\thetavec))^{-\gamma} \vert_{\theta_{j}=\theta_{j,m}}<\infty\) uniformly over \(\Delta\theta_{j}\). Recalling that \(\thetavec=[x,y],\) and assuming without loss of generality that \(\theta_{j}=x\), \(\Delta\theta_{j}=\Delta x\), we obtain 
\begin{align}\begin{array}{lll}
\left. D^{-\gamma/2}_{m}(\thetavec)\right\vert_{\theta_{j}=\theta_{j,m}}=D^{-\gamma/2}_{m}(x+\Delta x_{m},y)=\Dcond{\\[1em]~~~~~~~~~~~~~~~~~~~~~~}{}\left(\frac{1}{\sqrt{{(x+\Delta x_{m}-a_{m})^{2}+(y-b_{m})^{2}}}}\right)^{\gamma/2},
\end{array}\end{align} 
 where  \(\vert\Delta x_{m} \vert \leq \vert \Delta x\vert .\) After some algebraic manipulation and using the Bernoulli inequality\footnote{ \((1+x)^{r}\leq 1+r x\) for \(x>-1\) and \(r\in[0,1]\).},  
\begin{align}\label{Eq:Baound on Dm}\begin{array}{lll}
 D^{-1}_{m}(x+\Delta x_{m},y)\leq\frac{\left| \Delta x_m^2+2 a_m \Delta x_m\right| }{2 \left(a_m^2+b_m^2\right){}^{3/2}} +\frac{1}{\sqrt{a_m^2+b_m^2}}\leq\frac{5/2}{\sqrt{a_m^2+b_m^2}}
\end{array}\end{align}
where  the last inequality follows from the fact that for sufficiently large \(m,\)  $\sqrt{a_{m}^{2}+b^{2}_{m}}
>\vert\Delta x_m\vert.$ Since \eqref{Eq:Baound on Dm} implies  that \(D^{-1}_{m}(x+\Delta x_{m},y)\leq 5 D^{-1}_{m}(x,y)/2,\) uniformly on \(\Delta\theta_{j}\), and because \(\sum_{m=1}^{\infty}(D_{m}(\thetavec))^{-\gamma}<\infty\) it follows that \(\sum_{m=1}^{M_{0}}(D_{m}(\thetavec))^{-\gamma}\vert_{\theta_{j}=\theta_{j,m}}\leq \sum_{m=1}^{\infty}(D_{m}(\thetavec))^{-\gamma}\vert_{\theta_{j}=\theta_{j,m}}\leq \infty\) uniformly on \(\Delta\theta_{j}.\) Thus, \(\lim_{M_{1}\rightarrow\infty}\lim_{\Delta\theta_{j}\rightarrow\infty} V_{M_{1},\Delta\theta_{j}}=0\) in the m.s.   
It follows that 
 \beq
\lim_{\Delta\theta_{j}\rightarrow0}\frac{\Delta_{j}( \log p_{\thss\psiss}( \Rmat_{}))}{\Delta \theta_{j}}=\sum_{m=1}
^{\infty}\frac{\partial\log(p_{\thss,\psi_{m}}(R_{m}))
}{\partial\th_{j}}, m.s. 
\eeq
which establishes \eqref{derivativeUnderInfiniteSum}, thus establishing  
Proposition \ref{DerivativeOfTheLikelihood}.  \hfill $\blacksquare$ 
 
Back to the proof of Lemma \ref{Lemma7}; substituting \eqref{SimplifyDerivativeMarginal}, into \eqref{derivativeUnderInfiniteSum}   and then substituting the result  into   \eqref{FisherByDefinition}, it follows that 
\begin{align}
\begin{array}{lll}
[\Imat(\thb)]_{ij}=\E_{}\left\{\sum\limits_{m,n=1 }^\infty \frac{2}{N_{0}} \left.\intarr _{0}^\infty \frac{\partial s_{m}(t; \thetavec)} {\partial \theta_{i}}d W_{{m}}(t) \right. \Dcond{\right.\\~~~~~~~~~~~~~~~~~~~~~~~\times\left.}{}\frac{2}{N_{0}} \left.\intarr _{0}^\infty \frac{\partial s_{n}(t; \thetavec)} {\partial \theta_{j}}d W_{{n}}(t) \right.  \right\}\end{array}\end{align}
where the infinite sum and the expectation are interchangeable  (see e.g. \citep{Folland}, Proposition 5.21). Thus, 
\begin{align}
[\Imat(\bth)]_{ij}=\sum_{m,n=1}^{\infty}\delta_{m,n}\times[\ivec_{m}(\bth)]_{ij},\end{align}
where \(\delta_{m,n}\) is the Kronecker delta function, and \begin{align} \begin{array}{lll}[\ivec(\thb)]_{i, j}=\E\left\{\frac{2}{N_{0}} \left(\intarr _{0}^\infty \frac{\partial s_{m}(t; \thetavec)} {\partial \theta_{j}}d W_{_{m}}(t) \right)\Dcond{\right.\\~~~~~~~~~~~~~~~\times \left.}{} \frac{2}{N_{0}} \left(\intarr_{0}^\infty \frac{\partial s_{m}(t; \thetavec)}{\partial \theta_{i}}d W_{_{m}}(t )\right) \right\}\end{array}\\\label{Eq:FisherIntegralExpressionSecondLine} \mathop=\limits^{\rm }\frac{2}{N_{0}}\int_{0}^{\infty}\frac{ \partial s_{m}(t; \thetavec)} {\partial \theta_{j}}\frac{\partial s_{m}(t; \thetavec)} {\partial \theta_{i}}dt, \end{align}   where the second equality follows from the properties of the stochastic integral  (see e.g., \citep{poor1994introduction}, Proposition VI.D.1). This establishes \eqref{SumOfMarginalFIM}, thus establishing Lemma \ref{Lemma7}.  
 \hfill $\blacksquare$

\part{Derivation of the CRB  for the  model  in \eqref{e: system model}}\label{PartE} Now, using Proposition  \ref{Lemma7},
it is possible to derive  \(\Imat(\bth).\) We begin with   ${\partial s_{m}(t;\theta)}/{\partial\theta_{i}}$
for $i=1$ (recall that \(\theta_{1}=x\))
\beq\label{partial s_m partial x}\begin{array}{l}
\frac{\partial s_{m}(t)}{\partial x}=\frac{ \partial}{\partial
x}\left(k_0 D_{m}^{-\frac{\gamma }{2}} s\left(t-\frac{D_{m}}{c}\right)\right)=\\-\frac{k_0 }{2 c}D_m^{-\frac{\gamma }{2}-2} \left(x-\alpha _m\right) \left(2 D_m s'\left(t-\frac{D_m}{c}\right)+c \gamma  s\left(t-\frac{D_m}{c}\right)\right)
\end{array}\eeq
where we used  \({\partial D_m}/{\partial x}={(x-\alpha _m})/{D_m}\). The expression for $\partial s_{m}(t)/\partial
y$ can be obtained by substituting \(y\) and $\beta_{m}$
in $\partial s_{m}(t)/\partial x$, for \(x\) and $\alpha_{m}$,
respectively. Substituting \eqref{partial s_m partial x} into \eqref{Eq:FisherIntegralExpressionSecondLine}, one obtains
\beq [\ivec_{m}(\thetavec)]_{1,1}=\frac{E_{s}}{N_{0} } D_m^{-2} \left(x-\alpha _m\right){}^2 g(D_{m}),
\eeq
 where we used the assumption  \(\int _{0}^\infty s(t)\frac{ds(t)}{dt}dt=0\). The rest of
the entries  of \(\ivec_{m}(\thetavec)\) can be  
derived similarly to obtain
\begin{IEEEeqnarray}{ll}\label{}\ivec_{m}(\thetavec)=\rho
g(D_{m})\left[
\begin{array}{cc}
  \cos ^2 \phi _m  & \sin  \phi _m \cos  \phi _m  \\
 \sin  \phi _m\cos  \phi _m  &  \sin ^2 \phi _m
\end{array}
\right],
\end{IEEEeqnarray}
where  \(\cos \left(\phi _m\right)=\left(x-\alpha _m\right)/D_m\) and \(\sin \left(\phi _m\right)=\left(y-\beta _m\right)/D_m\).  Thus, the FIM is given by
\begin{IEEEeqnarray}{l}\label{e: FIM matrix} \Imat(\thetavec)
\\\nonumber=\rho \sum _{i=1}^M \left[
\begin{array}{cc}
{\cos ^2\left(\phi _i\right) g(D_i)} & \sin \left(\phi _i\right) \cos \left(\phi _i\right) g(D_i) \\
  \sin \left(\phi _i\right) \cos \left(\phi _i\right) g(D_i) &  \sin ^2\left(\phi _i\right) g(D_i)
\end{array}
\right]\end{IEEEeqnarray} The error \(\E\{\Vert\hat\btheta-\btheta\Vert^{2}\}=\E\{(\hat
x-x)^{2}+(\hat y-y)^{2}\}\}\)
is bounded by \beq\label{rawCRB}\E\{\Vert\hat\btheta-\btheta\Vert^{2}\}\geq\Tr(\Imat^{-1}
(\thetavec))=\frac{([\Imat(\btheta)]_{11}
+[\Imat(\thetavec)]_{22})}{\det(\Imat(\btheta))}\eeq
To obtain the bound, we first derive \(\det(\Imat(\thetavec))\),
\beqIl\nonumber \det(\Imat(\thetavec))&=\sum_{m=1}^{M}\sum_{j=1}^{M} [\ivec_{m}(\thetavec)]_{1,1}[\ivec_{j}(\btheta)]_{2,2}-
[\ivec_{m}(\btheta)]_{2,1}[\ivec_{j}(\btheta)]_{1,2}\\&=\rho^{2}\sum_{m=1}^{M}\sum_{
\mathop {n=m+1 }\limits_{}}^{M} g\left(D_m\right) g\left(D_n\right)   \sin^{2} \left(\phi _m-\phi _n\right)\eeqIl 
Substituting  the latter into \eqref{rawCRB} establishes the desired
result.\end{IEEEproof}

\section{Proof of Theorem \ref{Th:ExpectedCRBRaw}}
\label{ProofTheoremFiniteEnergy}
 We begin with some notation and definitions. Recall that sensor-locations  
 \(\bps=\{(a_{m},b_{m})\}_{m=1}^{\infty}\)
  is a realization of a homogeneous PPP  $\Psimat$ defined on \((\Omega,{\cal F},P)\); i.e., \(\bps=\bPs(\omega)\).
Denote by  \(c_{A}(\bpsi),\)  the number of points in the intersection of \(\bpsi\) with \(A\subset \bcal(\real^{2})\);
 i.e., \(c_{A}(\bps)=\sum_{m=1}^{\infty} \chi(\psi_{m}\in A)\), where \(\chi_{A}(\cdot)\) is the indicator function. Let \(C_{A}:\Omega\longrightarrow\real\) be the random variable \(C_{A}(\omega)=c_{A}(\Psimat(\omega)).\)  

Now to the proof. If  \eqref{FiniteSum} is satisfied, the measurability of $CRB(\thetavec,\bPs)$ follows immediately\footnote{ The bound  $CRB(\thetavec,\bPs)$ is a well defined random variable if it is   \((\Omega,\Fcal)-(\real,\Bcal(\real))\) measurable. Because the process \(\Psimat\) is a function  \(\Psimat:\Omega\longrightarrow {\Lambda}\) which is  \((\Omega,\Fcal)-(\Lambda,{\cal L})\) measurable, where \(\Lambda \) is the set of all locally finite sets in \(\real^{2}\); i.e., \(\psivec\in \Lambda\), if for   every  \(B\in\Bcal(\real^{2})\) with a finite Lebesgue-measure \(c_{B}(\psivec)<\infty\) and     \(\cal L\) is the minimal sigma-algebra of sets in \(\Lambda\) such that \(\psivec\mapsto c_{B}(\psivec)\) is measurable for every \(B\in\Bcal(\real^{2})\) with finite Lebesgue measure. Hence, to show that \(CRB(\thetavec,\Psimat)\) is a well defined random variable, it is sufficient to show that it is   \((\Lambda,{\cal L})-(\real,\Bcal(\real))\) measurable.} from the fact that its numerator and denominator, being a limit of measurable functions, are measurable functions of $\psivec$. It remains to show that \eqref{FiniteSum} is satisfied with probability 1.   Denote   \(B_{m}=\{\zvec\in\real^{2}:m\leq\Vert\zvec- \bth\Vert_{}<m+1\}\) for every \(m=0,1,2,...\) and    \(M_{{0}}=C_{B_{0}}\).  Then, for every \(M\in\nat\)   \beq \label{SumDmDecomposed}\sum_{ m=1}^{M}D_{m}^{-\gamma} \leq\underbrace {\sum\limits_{m = 1}^{M_{0}} D^{-\gamma}_{m} }_{{Q_{M_{0}}}} + \underbrace {\sum\limits_{m =1 }^M  {{m^{ - \gamma }}} C_{ B_{m}}}_{{K_{M}}};\eeq where  \(\sum_{m=M_{1}}^{M_{2}}z_{m}=0\) if \(M_{1}>M_{2}\). Thus, to show that \(\sum_{m=1}^{\infty} D_{m}<\infty\) it is sufficient to show that  each of the series  \(Q_{M_{0}}\) and \(K_{N}\) converges  \(P-\) a.s.  to some random variable\footnote{We use the standard definition for a   random variable; i.e.,   a Borel-measurable function \(X:\Omega\to\real\), where \(\real\) does not include \(\pm\infty\);  therefore,    \(\vert X(\omega)\vert<\infty\) for every \(\omega\).  Under this definition, if the limit diverges; i.e., \(\sum_{m=1}^{\infty}D_{M}=\infty\),  is not a random variable.}. Because  in homogenous PPP \(C_{ B_{m}}<\infty\) 
\(P-\)a.s. for every \(A\in\real^{2}\) with a finite Lebesgue measure, it follows that $M_{0}<\infty$ \(P-\) a.s. Thus, \(Q_{M_{0}}\) is a finite sum of random variables \(D^{-\gamma}_{1},...,D^{ -\gamma}_{M_{0}}\), where each \(D_{m}, m=1,...,M_{0}\) is a continuous random variable  \(D_{m}:\Omega\to[0,1)\). Because \(P(D_{m}=0)=0\), it is possible to define a continuous random variable \(Z_{m}:\Omega\to\real_{+}\) as  \(Z_{m}=D_{m}^{-\gamma}\) if \(D_{m}\neq 0\) and \(Z_{m}=0\) for \(D_{m}=0\). While \(\E\{ Z_{m}\}\) does not exist\footnote{The integral
diverges to infinity for every \(m\);
this is denoted
by \(\E\{Z_{m}\}=\infty\). If one considers the extended real line,  \(\bar \real=\real\cup\{-\infty,+\infty\}\)   the expectation is well defined.
In this paper we do not use \(\bar R\).}, \(Z_{m}\) is finite w.p. 1 and because \(M_{0}\) is finite w.p. 1, it follows that \(Q_{M_{0}}\) is a well defined random variable, and therefore finite. 

Next we show that \(K_{M}\) converges \(P-\) a.s. To this end, we use  the Khinchine Kolmogorov 1-series theorem (see e.g. \citep{athreya2006measure} Theorem 8.3.4)
\begin{theorem}[Khinchine-Kolmogorov's 1-series theorem]\label{3-seriesTheorem} Let \(\{X_{n}\}_{n=1}^{\infty}\)  be a sequence of independent random variables on a probability space \((\Omega,\fcal,P)\) such that \(\E\{X_{n}\}=0\) and \(\sum_{n=1}^\infty\E\{X_{n}^{2}\}<\infty\). Then,  \(I_{n}\define\sum_{j=1}^{n}X_{j}\) converges in the mean square and  \(P-\)
a.s.  as \(n\to\infty\).
\end{theorem}
 
Note that \(K_{M}\) is a weighted   sum of Poisson random variables  \(C_{ B_{m}}\) with mean and variance equal to \(v_{L}(B_{m})\lambda, \) where  \(v_{L}\) is the Lebesgue   measure on \(\real^{2}\). Therefore,    \(\E\{C_{ B_{m}}\}={\rm Var}(C_{ B_{m}}))=\lambda\pi (2m+1)\) and because
 \beqIl
\sum_{m =1}^{\infty}\E(m^{-\gamma}C_{ B_{m}}\Dcond{-&&}{-}\E\{m^{-\gamma}C_{ B_{m}}\})^{2}=
\sum_{m=1}^{\infty}{\rm Var}(m^{-\gamma}C_{ B_{m}})\Dcond{\nonumber\\&&}{}=\sum_{m=1}
^{\infty}m^{-2\gamma}\lambda\pi(2m+1)<\infty ,\eeqIl
 it follows that  
 \beq
 K'_{M}=\sum_{m=1}^{M}(m^{-\gamma}C_{ B_{m}}-m^{-\gamma}\E\{C_{ B_{m}}\})
 \eeq
converges \(P-\) a.s. Furthermore, note that 
\(\lim_{M\to\infty}\E\{K_{M}\}=\sum_{m=1}^{\infty}m^{-\gamma}\E\{ C_{B_{m}}\}=\pi\lambda\sum_{m=1}^{\infty}m^{-\gamma}(2m+1)<\infty
\)
%  \beqIl
%\lim_{M\to\infty}\E\{K_{M}\}\Dcond{&&}{}=\sum_{m=1}^{\infty}m^{-\gamma}\E\{ %C_{B_{m}}\}\Dcond{\\&&}{}=\pi\lambda\sum_{m=1}^{\infty}m^{-\gamma}(2m+1)<\infty
%\eeqIl 
where we replaced the order of the  expectation  with the infinite sum since \( C_{B_{m}}\) is  non-negative for every \(m\) (see e.g., \citep{Folland} Theorem 2.15). Hence  
\(\lim_{M\to\infty} K'_{M}+\lim_{M\to\infty}\E\{K_{M}\}=\lim_{M\to\infty} (K'_{M}+\E\{K_{M}\})=\lim_{M\to\infty} K_{M};\) 
i.e., \(K_{M}\) converges w.p. 1 to a random variable, and therefore, \(\lim_{M\to\infty} K_{M}<\infty\) \(P-\) a.s.

\bibliographystyle{ieeetr}
\bibliography{ISF2014}
\end{document}